\documentclass[12pt]{article} 

\usepackage{graphicx}
\usepackage{amssymb}
\usepackage{amsmath,amsfonts}
\usepackage{slashed}
\usepackage{mathtools}
\usepackage{enumerate}
\usepackage{accents}
\usepackage{titlesec}
\usepackage[title]{appendix}

\titleformat*{\section}{\large\bfseries\sffamily}

\newcommand{\IM} {\text{Im}\,}
\newcommand{\RE} {\text{Re}\,}
\newcommand{\Wr}{ {\cal W} }
\newcommand{\dbtilde}[1]{\accentset{\approx}{#1}}

\begin{document}

\normalsize

\begin{center}
\Large \bf Vortices and Factorization
\end{center}

\vspace{2mm}

\begin{center}
Igor Loutsenko and Oksana Yermolayeva
\end{center}

%
%\vspace{2mm}

\begin{center}
Laboratoire de Physique Math\'ematique,
CRM, Universit\'e de Montr\'eal
\end{center}

\vspace{5mm}

\begin{abstract}

{

We review applications of factorization methods to the problem of finding stationary point vortex patterns in two-dimensional fluid mechanics. Then we present a new class of patterns related to  periodic analogs of Schrodinger operators from the ``even" bispectral family. We also show that patterns related to soliton solutions of the KdV hierarchy constitute complete solution of the problem for certain classes of vortex systems.
 
}

\end{abstract}


%{\small \tableofcontents}

\section{Introduction, Stationary Vortex Configurations}
\label{vortex}

In fluid mechanics, factorization methods have found applications in the theory of two-dimensional flows in two opposite limits: the infinite-viscosity limit of the free-boundary flows (see, e.g., \cite{L,LY}), and the zero-viscosity limit. In this review we focus on the latter, namely applications to the problem of finding classes of two-dimensional stationary vortex configurations in inviscid two-dimensional flows.

Systems of point vortices are weak solutions of the two-dimensional Euler equation. Work on their classification began more than a century and a half ago and has been  substantially developed over the past several decades (for a general overview see, e.g., \cite{Ar1, Ar2, ANSTV, Cl, ON1, ON2}).

Motion of two-dimensional inviscid and incompressible liquid is described by the following system of PDEs
\begin{equation}
\frac{\partial \mathbf{v}}{\partial t}+(\mathbf{v}\cdot\nabla)\mathbf{v}=-\frac{\nabla P}{\varrho}, \quad \varrho={\rm const},
\label{Euler}
\end{equation}
\begin{equation}
\nabla \cdot \mathbf{v}=0,
\label{continuity}
\end{equation}
where ${\mathbf v}=\mathbf{v}(x,y,t)$, $\mathbf{v}=(v_x,v_y)$ is the two-dimensional velocity of the flow, and $P=P(x,y,t)$ stands for the pressure. Equation (\ref{Euler}) is the Euler's equation, while (\ref{continuity}) is the continuity equation. In two dimensions, the vorticity, which is the curl of the velocity field, has a single component $\omega$:
$$
\omega=\frac{\partial v_y}{\partial x}-\frac{\partial v_x}{\partial y} .
$$
Taking the curl of (\ref{Euler}) we get
$$
\frac{d\omega}{dt}=\frac{\partial\omega}{\partial t}+(\mathbf{v}\cdot\nabla)\omega=0.
$$
In other words, the total derivative of the vorticity along the flow is zero. It follows that the circulation $\int_{\Omega(t)} \omega dxdy=\oint_{\partial\Omega(t)}v_xdx+v_ydy$ along the boundary of any evolving with the flow domain $\Omega(t)$ does not change with time. As a consequence, for the system of point vortices
$$
\omega=\sum_{i=1}^N {\cal Q}_i\delta(x-x_i(t))\delta(y-y_i(t))
$$
the strength (i.e., circulation) of each vortex ${\cal Q}_i$ is time independent, and the speed of the $i$th vortex $(dx_i/dt,dy_i/dt)$ equals the velocity of the flow averaged over a small circular domain centered at $(x_i,y_i)$. Since vorticity is the curl of the velocity, for the system of point vortices we have
\begin{equation}
v_x-{\rm i}v_y=V_x-{\rm i}V_y+\frac{1}{2\pi {\rm i}}\sum_{i=1}^N\frac{{\cal Q}_i}{z-z_i(t)} ,
\label{uw}
\end{equation}
where we use complex notations $z=x+{\rm i}y$. In (\ref{uw}), the full velocity field $\mathbf{v}=(v_x,v_y)$ is the sum of the irrotational (i.e., zero-curl) background flow $\mathbf{V}=(V_x,V_y)$ and the flow created by superposition of $N$ point vortices. From (\ref{continuity}, \ref{uw}), $\mathbf{V}$ is a zero-divergence field. Since it is irrotational, it is also a gradient of a potential, and consequently, the potential is a harmonic function. In Sections \ref{vortex}-\ref{translating}, we will consider unrestricted flows in the whole plane with velocities bounded at infinity. The only harmonic functions satisfying above conditions are linear functions, so the background flow has constant velocity (uniform background flow). Using (\ref{uw}), we find the average velocity of the flow in a small disc centered at $z=z_i$, thus, obtaining the following equations of motion of vortices in the uniform background flow
\begin{equation}
2\pi {\rm i} \frac{d\bar z_i}{dt}=k+\sum_{j=1, j\not=i}^N\frac{{\cal Q}_j}{z_i-z_j}, \quad i=1\dots N,
\label{vmotion}
\end{equation}
where $k$ is a complex constant, and the overbar denotes the complex conjugation.

The classification of the stationary patterns of vortices is then reduced to the study of solution of the system of $N$ algebraic equations
\begin{equation}
k+\sum_{j=1, j\not=i}^N\frac{{\cal Q}_j}{z_i-z_j}=0, \quad i=1\dots N .
\label{stationary}
\end{equation}
This system also has a two-dimensional electrostatic interpretation: (\ref{stationary}) is nothing but an equilibrium condition for $N$ point electric charges ${\cal Q}_1, \dots, {\cal Q}_N$ interacting pairwise through a two-dimensional Coulomb (logarithmic) potential, and placed in the homogeneous electric field $\bar k$. The equilibrium configuration is a stationary point of the electrostatic energy
$$
E=k\sum_i {\cal Q}_i z_i + \sum_{i<j} {\cal Q}_j{\cal Q}_j \log(z_i-z_j) + {\rm c.c.} ,
$$
where ``c.c." stands for the complex conjugate.

The stationary patterns with $k\not=0$ are called translating equilibria (or translating configurations/patterns) because in the reference frame where the fluid rests at infinity, the whole pattern translates with constant velocity $(\IM k,\RE k)/(2\pi)$. Below, we mainly deal with equilibria without background flow, i.e., $k=0$ equilibrium configurations (``static configurations/patterns"):
\begin{equation}
\sum_{j=1, j\not=i}^N\frac{{\cal Q}_j}{z_i-z_j}=0, \quad i=1\dots N .
\label{static}
\end{equation}
This system of equations is invariant under rigid motions and scaling of the plane. From the invariance of the stationary value of the electrostatic energy $E=2\sum_{i<j} {\cal Q}_i{\cal Q}_j\log|z_i-z_j|$ under scaling of the plane ($z_i\to c z_i$), it follows that the system has solutions only if
\begin{equation}
\sum_{i<j}{\cal Q}_i{\cal Q}_j=0 .
\label{scaling}
\end{equation}
O'Neil showed \cite{ON1} that for almost all values of $Q_i$, satisfying the above condition, the number of distinct solutions to (\ref{static}), modulo rotations, translations and scaling of the plane, is $(N-2)!$.

In the case when $k\not=0$ the translational invariance of the stationary value of energy $E=\sum_i {\cal Q}_i(kz_i+\bar k\bar z_i)+2\sum_{i<j} {\cal Q}_i{\cal Q}_j\log|z_i-z_j|$ leads to the following (``neutrality") condition
\begin{equation}
\sum_i {\cal Q}_i=0 .
\label{neutrality}
\end{equation}
For almost all values of ${\cal Q}_i$ satisfying this condition, the number of distinct, modulo translations of the plane, solutions of (\ref{stationary}) is $(N-1)!$ \cite{ON1}.

In other words, for generic ${\cal Q}_i$, satisfying (\ref{scaling}) or (\ref{neutrality}), dimension of the solution set is zero. There are, however, ``resonant" values of ${\cal Q}_i$, where this dimension is positive, i.e., solutions depend on non-trivial continuous complex parameters. This paper mainly deals with such resonant cases. Sections \ref{streets} and \ref{streetseven} are devoted to periodic (``vortex street") versions of such systems.

We also briefly consider non-resonant cases. These are equilibrium patterns in non-uniform background flows, discussed in Section \ref{nonuniform}, or a class of periodic patterns, discussed at the beginning of Section \ref{streetseven} . Such patterns are generated by applying Darboux transformations to the ``classical" configurations, which were previously obtained using the polynomial method \cite{Cl,Lbhe}. The ``classical" configurations, first found  by Stieltjes \cite{St}, are related to classical orthogonal polynomials, such as Hermite, Laguerre, and Jacobi polynomials (see, e.g., \cite{Ar2,MMM})\footnote{Among the other configurations analyzed using the polynomial method, one can mention nested vortex polygons in rotating fluids \cite{Ar2,ANSTV}. We do not discuss these here.}.

The polynomial method is a technique that uses polynomial solutions to differential equations to find and classify stationary configurations of point vortices. Specifically, vortex positions are linked to the roots of certain polynomials, enabling the study of equilibria and their connections to various polynomial systems.  

Significant advancements in the field followed the works of Tkachenko \cite{Tk} and Bartman \cite{Bar}. In particular, Bartman observed connections between the equilibria of systems consisting of vortices of two species with a circulation ratio $-1$ and rational solutions of the KdV hierarchy.

In the next two sections, we review the application of the polynomial method to systems consisting of finite number of vortices of two species with the circulation ratio $-1$, or circulation ratio $-2$. For such systems, the polynomial method allows one to demonstrate, through a recursive procedure, the existence of infinite sequences of static configurations with increasing dimensions of the solution set \cite{Bar,BC,Lv}. These sequences exhaust all possible static configurations for such systems.

In the case of the circulation ratio $-1$, the recursive procedure can be reformulated in terms of transformations of the Schrodinger operator through factorization (Darboux transformations). Then, factorization method enables one to generate all possible translating configurations as well as to find the Wronskian representation of configurations. In more general setting, it enables the construction of families of vortex street patterns, and families of configurations in the non-uniform background flows. In cases where configurations are related to rational or soliton solutions of the KdV hierarchy ($\Lambda=1$ configurations of Section \ref{L12}, and  those of sections \ref{translating}, \ref{streets}) there exist alternatives to factorization method, e.g. $\tau$-function method.

In the case of the circulation ratio $-2$, static configurations can be generated through factorization of the third-order operators (considered in Sections \ref{3rd_order} and \ref{3rd_order_equilibrium}) \cite{LY1}. These configurations are related to rational solutions of the Sawada-Kotera hierarchy, and their Pfaffian representation can be obtained using the $\tau$-function method (Appendix \ref{AA}).

We also review families of multi-parametric configurations of three species of vortices (discussed in Sections \ref{Bispectral}, \ref{KWCCterminating}, and \ref{3rd_order_equilibrium}). These families, initially found using polynomial method\footnote{Actually, one of these families was first constructed through factorization, but in a different context \cite{DG}.}  \cite{ONCS}, can be constructed through the factorization of either the Schrodinger or third-order operators. A new (vortex-street) generalization of one such family is obtained using the factorization method in Section \ref{streetseven}.

\section{Vortices, Polynomials and Rational Primitives}
\label{polynomials}

Let us first consider static equilibria of $N=l+m$ charges (or vortices) of two species with a circulation ratio $-\Lambda$. Without loss of generality we can set the values of the first $l$ charges to $-1$ and those of the remaining $m$ charges to $\Lambda$:
$$
{\cal Q}_i=\left\{
\begin{array}{ll}
-1, & i=1\dots l\\
\Lambda, & i=l+1 \dots l+m
\end{array}
\right. .
$$
Next, we introduce the following polynomials in $z$
\begin{equation}
p(z)=\prod_{i=1}^l (z-z_i), \quad q(z)=\prod_{i=1}^m (z-z_{l+i}) ,
\label{pzqz}
\end{equation}
whose roots correspond to positions of charges of the first and second species respectively. Since we are dealing with two distinct species, $p$ and $q$ do not have common or multiple roots (For cases involving common/multiple roots, see e.g. \cite{DK2}). The static equilibrium condition (\ref{static}) for this system of two species is equivalent to the following bi-linear differential equation for the above polynomials
\begin{equation}
p''q-2\Lambda p'q'+\Lambda^2pq''=0,
\label{bilinear}
\end{equation}
where prime denotes differentiation with respect to $z$. One can check this equivalence by applying partial fraction decomposition to the ratio of the left-hand side of (\ref{bilinear}) and $pq$: Using the identity $\frac{1}{(z-z_i)(z-z_j)}=\frac{1}{z_i-z_j}\left(\frac{1}{z-z_i}-\frac{1}{z-z_j}\right)$ and rearranging terms, we obtain the following expression for the ratio\footnote{For a more detailed derivation of poly-linear equations, see e.g. section 5.1 of \cite{ON2}.}
$$
\sum_i\frac{2{\cal Q}_i\sum_{j\not=i}\frac{{\cal Q}_j}{z_i-z_j}}{z-z_i}, \quad {\cal Q}_i\in\{-1,\Lambda\} .
$$
This expression vanishes identically in $z$ if and only if each residue $2{\cal Q}_i\sum_{j\not=i}\frac{{\cal Q}_j}{z_i-z_j}$ equals zero, i.e., if and only if $\sum_{j\not=i}\frac{{\cal Q}_j}{z_i-z_j}=0$. The latter is precisely the equilibrium conditions (\ref{static}). Therefore (\ref{bilinear}) and (\ref{static}) are equivalent when ${\cal Q}_i\in\{-1,\Lambda\}$.

The $\Lambda=1$ specification of (\ref{bilinear})
\begin{equation}
p''q-2 p'q'+pq''=0, 
\label{Tkachenko}
\end{equation}
is called the Tkachenko equation \cite{Tk} in the literature on vortex patterns. Its generic solutions (i.e., those without common/multiple roots) describe equilibria of two species of vortices (charges) of equal magnitude but opposite sign. Bartman \cite{Bar} identified the Tkachenko equation with the recurrence relation (\ref{AM}) for the Adler-Moser polynomials. As shown in the works of Burchnall and Chaundy \cite{BC} (see the next section), any generic polynomial solution of the Tkachenko equation is (modulo multiplication by arbitrary constants, and a shift of $z$) a pair of the Adler-Moser polynomials \cite{AM}:
\begin{equation}
p(z)=P_{n \pm 1}(z), \quad q(z)=P_n(z) .
\label{pqP}
\end{equation}
In this way, we obtain a complete classification of static configuration of a system of vortices (charges) with strengths $\pm 1$. The degree of the $n$th Adler-Moser polynomial equals $n(n+1)/2$, which can be derived from general considerations without analyzing the Tkachenko equation. Indeed, static configurations exist only when the condition (\ref{scaling}) holds. For charges ${\cal Q}_i=\pm 1$, this condition is satisfied only if $l$ and $m$ are consecutive triangular numbers. Therefore, equilibrium is only possible when the number of negative and positive charges (or vice versa) equals $l=n(n-1)/2$ and $m=n(n+1)/2$. The equilibrium positions of charges are the roots of consecutive Adler-Moser polynomials. These polynomials possess a non-trivial property: $P_n(z)$ is also a function of $n$ free parameters\footnote{Including the shift of $z$. Without loss of generality, this trivial parameter can be set to zero and is excluded from the standard definition of Adler-Moser polynomials.}, so all possible equilibria of $N=l+m=n^2$ charges form an $n$-dimensional complex subspace of the $n^2$ dimensional complex space.

Details on Adler-Moser polynomials will be provided in the following sections. Here, we note that these polynomials were first found by Burchnall and Chaundy \cite{BC}, later re-discovered in the context of the Hadamard problem \cite{LS} by Lagnese and Stellmacher, and subsequently by Adler and Moser as polynomial $\tau$-functions corresponding to rational solutions of the KdV hierarchy of integrable PDEs \cite{AM}. Burchnall and Chaundy \cite{BC} constructed these polynomials as solutions of the Tkachenko equation (\ref{Tkachenko}) by considering it as a linear second order ODE for $p$, where coefficients of ODE are determined by some polynomial $P_n=q$. Then the requirement that its two linearly independent solutions $p=P_{n-1}$ and $p=P_{n+1}$ are both polynomial is equivalent to rationality in $z$ of two indefinite integrals (rational primitives)
$$
\int \frac{p^2(z)}{q^2(z)}dz, \quad \int \frac{q^2(z)}{p^2(z)}dz .
$$
It turned out that, in addition to the Adler-Moser ($\Lambda=1$) case, there is another case with $\Lambda=2$, where an infinite chain of polynomial solutions to (\ref{bilinear}) exists and can be constructed with the help of rational primitives \cite{Bar, Lv}.

\section{$\Lambda=1$ and $\Lambda=2$ Chains: Main Sequences}
\label{L12}

Let us return to the generic $\Lambda$ case, and consider the bi-linear equation (\ref{bilinear}) as a second order linear ODE for $p=p_{n-1}(z)$ with coefficients of the equation being defined by $q=q_n(z)$. By elementary methods we find that the second linearly independent solution $p=p_n(z)$ equals
$$
p_n(z)=C_n p_{n-1}(z)\int \frac{q_n(z)^{2\Lambda}}{p_{n-1}(z)^2} dz ,
$$
where $C_n$ is a constant\footnote{The values of such constants are typically chosen so that $p_n$ and $q_n$ are monic in $z$.}. In order to build a recursive chain we require that both $p_{n-1}$ and $p_n$ be polynomial in $z$, so the primitive
\begin{equation}
\int \frac{q^{2\Lambda}}{p^2} dz
\label{qp}
\end{equation}
must be rational. Alternatively, by considering (\ref{bilinear}) as an ODE for $q$, we arrive at the requirement of rationality of
\begin{equation}
\int \frac{p^{2/\Lambda}}{q^2} dz .
\label{pq}
\end{equation}
It is clear that the pair of the rationality conditions above requires that both $2\Lambda$ and $2/\Lambda$ be integers, which is possible only if
\begin{equation}
\Lambda \in \left\{\frac{1}{2}, 1, 2\right\} .
\label{Lambda}
\end{equation}
The case $\Lambda=1$ has already been mentioned and corresponds to the Adler-Moser polynomials. Since equation (\ref{bilinear}) is invariant under the involution $\Lambda \leftrightarrow 1/\Lambda$, $p \leftrightarrow q$, the cases $\Lambda=1/2$ and $\Lambda=2$ are equivalent. Therefore, without loss of generality, both can be represented by the $\Lambda=2$ case.

Now, let us take $p$ and $q$ that do not have multiple/common roots and factorize $p(z)$ as $p(z)=(z-z_i)\tilde p(z)$, where $z_i, i\le l$ is a root of $p$ (see definition (\ref{pzqz})). Then, residue of a simple pole at $z=z_i$ in (\ref{qp}) equals
$$
\frac{\partial}{\partial z}\left(\frac{q(z)^{2\Lambda}}{\tilde p(z)^2}\right)_{z=z_i}=\frac{q(z_i)^{2\Lambda-1}}{\tilde p(z_i)^3}(2\Lambda\tilde p(z_i)q'(z_i)-2\tilde p'(z_i)q(z_i)) .
$$
Since $\tilde p(z_i)=p'(z_i)$ and $\tilde p'(z_i)=p''(z_i)/2 $, the expression for the residue becomes
$$
-\frac{q(z_i)^{2\Lambda-1}}{\tilde p(z_i)^3}\left(p''(z_i)q(z_i)-2\Lambda p'(z_i)q'(z_i)\right) .
$$
The last factor in the above expression vanishes, since $p$ and $q$ satisfy (\ref{bilinear}). Due to the absence of multiple roots, $\tilde p(z_i)\not= 0$, and the entire expression equals zero. Therefore, a simple pole is absent at $z=z_i$, and (\ref{qp}) is rational when (\ref{bilinear}) holds. By exchanging $p$ with $q$, $\Lambda$ with $1/\Lambda$, and repeating all the above arguments, we can demonstrate rationality of (\ref{pq}).

It is worth mentioning that the converse statement also holds, i.e. the bilinear equation (\ref{bilinear}) follows from the rationality of (\ref{qp}) and (\ref{pq}) (for details, see \cite{Lv}). 

Thus, once a pair  of polynomials $p,q$ without multiple or common roots satisfying (\ref{bilinear},\ref{Lambda}) is found, one can construct an infinite chain of polynomial solutions of (\ref{bilinear}) with the help of recursive procedure for $p_n$ and $q_n$ mentioned at the beginning of this section.

In more detail: Since $p=z^l+\dots $ and $q=z^m+\dots$, from the leading term of (\ref{bilinear}) we get the Diophantine equations\footnote{These equations can be also obtained from (\ref{scaling}).} connecting $l$ and $m$
\begin{itemize}

\item $\Lambda=1$ case: $(l-m)^2=l+m$,

\item $\Lambda=2$ case: $(l-2m)^2=l+4m$.

\end{itemize}
These are quadratic equations, and to each $l$ there correspond two values of $m$ and vice versa (since to each $p$ there correspond two $q$'s and vice versa). In other words, sequence of solutions has the form
$$
\dots , (l_i,m_i), (l_i,m_{i+1}), (l_{i+1},m_{i+1}), \dots
$$
where
\begin{itemize}
\item $\Lambda=1$ (Adler-Moser) case: $l$ and $m$ are consecutive triangular numbers
\begin{equation}
l_i=i(2i+1), \quad m_i=i(2i-1).
\label{lmlambda1}
\end{equation}
In terms of standard notations for the Adler-Moser polynomials $P_n$, we have $p_i=P_{2i}$ and $q_i=P_{2i+1}$, and (\ref{Tkachenko}) writes as
\begin{equation}
P_n''P_{n-1}-2P_n'P_{n-1}'+P_nP_{n-1}''=0, \quad \deg(P_n)=n(n+1)/2 .
\label{AM}
\end{equation}
\item $\Lambda=2$ case:
\begin{equation}
l_i=i(3i+2), \quad m_i=\frac{i(3i-1)}{2}.
\label{lmlambda2}
\end{equation}
\end{itemize}
Now, consider bilinear relation (\ref{bilinear}) with $p=p_i$ and $q=q_i$ as a linear second-order ODE with solution $q_i$. Its second linearly independent solution, $q=q_{i+1}$, is given by:
\begin{equation}
q_{i+1}=\left(\frac{2l_i}{\Lambda}-2m_i+1\right)q_i\int \frac{p_i^{2/\Lambda}}{q_i^2}dz .
\label{qup}
\end{equation}
Similarly, considering $p_{i-1}$ and $p_i$ as linearly independent solutions, we obtain:
\begin{equation}
p_i=\left(2\Lambda m_i-2l_{i-1}+1\right)p_{i-1}\int \frac{q_i^{2\Lambda}}{p_{i-1}^2}dz .
\label{pup}
\end{equation}
Due to freedom in choosing linearly independent solutions of (\ref{bilinear}), one can also write analogs of (\ref{qup}) and (\ref{pup}) for decreasing indices:
\begin{equation}
q_i=\left(\frac{2l_i}{\Lambda}-2m_{i+1}+1\right)q_{i+1}\int \frac{p_i^{2/\Lambda}}{q_{i+1}^2}dz ,
\quad p_{i-1}=\left(2\Lambda m_i-2l_i+1\right)p_i\int \frac{q_i^{2\Lambda}}{p_i^2}dz .
\label{qpdown}
\end{equation}
Thus, one can generate $p_i$, and $q_i$ iteratively in either directions starting at some $i$.

Rewriting (\ref{qup},\ref{pup}) or (\ref{qpdown}) in the differential form, we obtain the first-order differential recurrence relations:
\begin{equation}
\begin{array}{ll}
q'_{i+1}q_i-q_{i+1}q'_i=\left(\frac{2l_i}{\Lambda}-2m_i+1\right)p_i^{2/\Lambda}, \\
p'_ip_{i-1}-p_ip'_{i-1}=\left(2\Lambda m_i-2l_{i-1}+1\right)q_i^{2\Lambda}
\end{array}.
\label{Bartman}
\end{equation}
In the case $\Lambda=1$, the above recurrence relations become the first-order relations for the Adler-Moser polynomials $P_n$, with $q_i=P_{2i}$ and $p_i=P_{2i+1}$:
\begin{equation}
P'_{i+1}P_{i-1}-P'_{i-1}P_{i+1}=(2i+1)P^2_i .
\label{AM3}
\end{equation}
Examples of several first Adler-Moser polynomials are :
\begin{equation}
P_0=1, \quad P_1=z, \quad P_2=z^3+s_1, \quad P_3=z^6+5s_1z^3+s_2z-5s_1^2, \quad \dots ,
\label{AMExamples}
\end{equation}
where $s_i$ are free parameters related bi-rationally to the KdV ``times" \cite{AM, AcV} (we set parameter related to shift of $z$ to zero). For more details on this relationship, see Appendix \ref{AA}.

In the case $\Lambda=2$, we obtain the following recurrence relations:
\begin{equation}
\begin{array}{ll}
q'_{i+1}q_i-q_{i+1}q'_i=(3i+1)p_i, \\
p'_ip_{i-1}-p_ip'_{i-1}=\left(6i-1\right)q_i^4,
\end{array}
\label{qqp_ppq}
\end{equation}
where we used (\ref{lmlambda2}). Unlike the Adler-Moser case, there are two distinct branches of polynomials, both starting at $p_0=1$ and $q_0=1$. One branch, with $i\ge 0$, goes in the positive direction:
\begin{equation}
\begin{array}{ll}
q_0=1 & \quad p_0=1\\
q_1=z & \quad p_1=z^5+r_1\\
q_2=z^5+s_2z-4r_1 &
\begin{array}{r}
\;\; p_2={z}^{16}+{\frac {44}{7}}s_{{2}}{z}^{12}-32r_{{1}}{z}^{11}+22{s_{{2}}}^{2}{z}^{8}
-{\frac {2112}{7}}r_{{1}}s_{{2}}{z}^{7}\\+
1408{r_{{1}}}^{2}{z}^{6}+r_{{2}}{z}^{5}
-44{s_{{2}}}^{3}{z}^{4}+352r_{{1}}{s_{{2}}}^{2}{z}^{3}\\-1408s_{{2}}{r_{{1}}}^{2}{z}^{2}+
2816{r_{{1}}}^{3}z+r_{{2}}r_{{1}}-{\frac {11}{5}}{s_{{2}}}^{4}
\end{array}
\\
\dots & \quad \dots \\
q_n=q_n(z;r_1,r_2, \dots r_{n-1}; s_2, s_3, \dots s_n) & \quad p_n=p_n(z; r_1,r_2, \dots, r_n; s_2, s_3, \dots s_n) \\
\dots & \quad \dots
\end{array}
\label{pqplus}
\end{equation}
etc, and another branch, with $i\le 0$, goes in the negative direction:
\begin{equation}
\begin{array}{ll}
p_0=1 & \quad q_0=1 \\
p_{-1}=z & \quad q_{-1}=z^2+s_{-1} \\
\begin{array}{r}
p_{-2}=z^8+\frac{28}{5}s_{-1}z^6+14s_{-1}^2z^4\\
+28 s_{-1}^3z^2+r_{-2}z-7 s_{-1}^4
\end{array}
& \quad
\begin{array}{r}
q_{-2}=z^7+7 s_{-1}z^5+35 s_{-1}^2z^3+ s_{-2}z^2\\
-35 s_{-1}^3z+ s_{-1} s_{-2}-\frac{5}{2}r_{-2}
\end{array}\\
\dots & \quad \dots \\
p_{-n}=p_{-n}(z; r_{-2},r_{-3} \dots r_{-n}; s_{-1},s_{-2}, \dots s_{1-n}) & \quad q_{-n}=q_{-n}(z;r_{-2},r_{-3} \dots r_{-n}; s_{-1},s_{-2}, \dots s_{-n}) \\
\dots & \quad \dots
\end{array}
\label{pqminus}
\end{equation}
etc, where $r_i$ and $s_i$ are arbitrary complex constants \footnote{Without loss of generality we omit a constant corresponding to a shift of $z$, so that $q_1=p_{-1}=z$.} emerging in the course of integrations in (\ref{qup},\ref{pup}) or (\ref{qpdown}). Here, we use the following normalization for the constants:
\begin{equation}
q_{\pm n}=\tilde q_{\pm n}+s_{\pm n}q_{\pm(n-1)}, \quad p_{\pm n}=\tilde p_{\pm n}+r_{\pm n}p_{\pm(n-1)},
\label{sr}
\end{equation}
where the term proportional to $z^{\deg q_{\pm(n-1)}}$ is absent in $\tilde q_{\pm n}$ and the term proportional to $z^{\deg p_{\pm(n-1)}}$ is absent in $\tilde p_{\pm n}$. 

Summarizing, we emphasize again that for $\Lambda=1$ and $\Lambda=2$, any solutions of the bi-linear equation (\ref{bilinear}) with no common/multiple roots can be obtained through the recursive procedure described in this section. Therefore, the main sequences (\ref{AMExamples}) or (\ref{pqminus},\ref{pqplus}) determine all possible static patterns of a finite number of vortices with ${\cal Q}_i\in \{-1, 1\}$ or ${\cal Q}_i\in \{-1,2\}$, respectively.

\section{Darboux Transformations and Factorization Chains}
\label{DTFC}

The results above can also be obtained using factorization method. This approach not only recovers the previous results, but also enables the construction of new families of stationary configurations. In this section, we review the factorization method for Schrodinger operators. Third-order operators will be considered in Sections \ref{3rd_order} and \ref{3rd_order_equilibrium}.

The Schrodinger operator (``quantum Hamiltonian") with potential $u(z)$ has the form:
$$
H=-\partial_z^2+u(z) .
$$
We can express it as
\begin{equation}
H=A^*A+\eta,
\label{HAA}
\end{equation}
where $\eta$ is a constant and $A$, $A^*$ are formally adjoint first-order operators\footnote{Note that we do not use the special symbol $\circ$ for the composition of operators. Instead, we write expressions like $A^*A$ or $\varkappa\partial_z\varkappa^{-1}$, without using the composition symbol.}:
\begin{equation}
A^*=-\varkappa^{-1}\partial_z\varkappa=-\partial_z-\varkappa'/\varkappa, \quad A=\varkappa\partial_z\varkappa^{-1}=\partial_z-\varkappa'/\varkappa.
\label{AstarA}
\end{equation}
The function $\varkappa=\varkappa(z)$ is an eigenfunction of $H$ corresponding to the eigenvalue $\eta$
$$
H\varkappa=\eta\varkappa .
$$
By permuting factors $A^*$ and $A$ in (\ref{HAA}), we get the new Schrodinger operator $\hat H$
\begin{equation}
\hat H=AA^*+\eta=-\partial_z^2+\hat u, \quad \hat u=u-2(\log\varkappa)'' .
\label{DTH}
\end{equation}
From (\ref{HAA}) and (\ref{DTH}) it follows that for any constant $\lambda$
\begin{equation}
A(H-\lambda)=(\hat H-\lambda)A .
\label{HlambdaAAHlambda}
\end{equation}
Now, let us take an eigenfunction $\psi$ corresponding to the eigenvalue $\lambda$ of the original Schrodinger operator, i.e. $H\psi=\lambda\psi$. Then, from (\ref{HlambdaAAHlambda}) we see that following transform of $\psi$:
\begin{equation}
\hat \psi=A \psi=\psi'-\frac{\varkappa'}{\varkappa}\psi,
\label{DTpsi}
\end{equation}
is an eigenfunction of the new operator $\hat H$ corresponding to the same eigenvalue $\lambda$, i.e.,
$$
\hat H\hat\psi=\lambda\hat\psi .
$$
In other words, we can obtain a new operator $\hat H$ and its eigenfunctions $\hat \psi=\hat\psi(z;\lambda)$ from the old operator $H$ and its eigenfunctions $\psi=\psi(z;\lambda)$ corresponding to the same eigenvalues $\lambda$. New operator and eigenfunctions are obtained with the help of a ``seed" eigenfunction $\varkappa=\psi(z;\eta)$ of the old operator $H$, corresponding to the ``seed" eigenvalue $\eta$. The transformation $H\to\hat H$, $\psi\to\hat\psi$ given by (\ref{DTH},\ref{DTpsi}) is called the Darboux Transformation.

For all eigenvalues $\lambda$, except the seed eigenvalue, i.e. for $\lambda\not=\eta$, the transform (\ref{DTpsi}) of the two dimensional eigenspace of $H$ is also two-dimensional. However, when $\lambda=\eta$, the transformation (\ref{DTpsi}) annihilates the eigenfunction $\varkappa=\psi(z;\eta)$ of $H$, i.e. $A\varkappa=0$: It maps the two-dimensional kernel $C_1\varkappa+C_2\varkappa\int\frac{dz}{\varkappa^2} $ of $H-\eta$ to the one dimensional sub-space
$$
A\left[C_1\varkappa+C_2\varkappa\int\frac{dz}{\varkappa^2}\right]=\frac{C_2}{\varkappa}.
$$
From the above, we see that function $1/\varkappa$ is an eigenfunction of the new Schrodinger operator corresponding to the eigenvalue $\eta$:
$$
\hat H\varkappa^{-1}=\eta\varkappa^{-1} .
$$
Then, the general solution $\hat\varkappa$ of $\hat H\hat\varkappa=\eta\hat\varkappa$ is
\begin{equation}
\hat \varkappa = \frac{C}{\varkappa}\int \varkappa^2 dz, 
\label{DT0}
\end{equation}
where $C$ is an arbitrary constant and the primitive of $\varkappa^2$ includes another arbitrary constant of integration. This extends transformation of an eigenfunction (\ref{DTpsi}) to the case $\lambda=\eta$. Iterations of transformation (\ref{DT0}) are typically considered for $\eta=0$. In such cases, transformations (\ref{DT0}) are reffered to as Darboux transformations at the zero energy level.

In general, the sequence of iterated transformations is presented by the factorization chain, usually referred to as the chain of Darboux transformations at levels \footnote{It would be more accurate to refer to it as the chain of Darboux transformations {\it associated with} levels $\lambda=\eta_i$, rather than {\it at} levels  $\lambda=\eta_i$, since eigenfunctions at all values of $\lambda$ are transformed.} $\lambda=\eta_i$:
\begin{equation}
H_0=A_0^* A_0+\eta_0 \to H_1=A_0A_0^*+\eta_0=A_1^* A_1+\eta_1 \to H_2=A_1A_1^*+\eta_1=A_2^* A_2+\eta_2 \to \dots .
\label{dtchain}
\end{equation}
When all $\eta_i=0$, (\ref{dtchain}) becomes the chain of transformations at the zero energy level:
\begin{equation}
H_0=A_0^* A_0 \to H_1=A_0A_0^*=A_1^* A_1 \to H_2=A_1A_1^*=A_2^* A_2 \to \dots ,
\label{rchain}
\end{equation}
where
$$
A_i=\varkappa_i\partial_z\varkappa_i^{-1}, \quad A_i^*=-\varkappa_i^{-1}\partial_z\varkappa_i .
$$
According to (\ref{DT0}) and (\ref{DTH}),
\begin{equation}
\varkappa_{i+1}=\frac{C_i}{\varkappa_i}\int \varkappa_i^2 dz, \quad u_{i+1}=u_i-2(\log\varkappa_i)'', \quad H_i\varkappa_i=0,
\label{muj}
\end{equation}
where $C_i$ are arbitrary multiplication constants, and the primitives include arbitrary constants of integration. Using (\ref{muj}), we can construct iteratively the chain (\ref{rchain}). At each step of the chain a free parameter (arising from the constant of integration) appears  in (\ref{muj}).

It follows from (\ref{dtchain}) that $H_0$ and $H_n$ are related by the intertwining operator\footnote{Additionally, conjugate (``inverse") intertwining identities hold: $T^*_nH_n=H_0T^*_n$, $T^*_n=A^*_0 \cdots A^*_{n-2}A^*_{n-1}$.} $T_n$:
\begin{equation}
H_nT_n=T_nH_0, \quad T_n=A_{n-1}A_{n-2}\cdots A_0 .
\label{HTTH}
\end{equation}
The function $T_n\psi(x;\lambda)$, where $\psi$ is an eigenfunction of $H_0$, is an eigenfunction of $H_n$. According to the Crum therem \cite{Crum}, the intertwining operator as well as potential $u_n$ can be explicitly written down in terms of eigenfunctions of the initial hamiltonian $H_0$. In more detail, the theorem states that the composition of $n$ Darboux transformations, corresponding to the factorization chain (\ref{dtchain}), transforms the eigenfunction $\psi$, corresponding to the eigenvalue $\lambda$ of $H_0=-\partial_z^2+u_0$ to the eigenfunction $\hat\psi$ of $H_n=-\partial_z^2+u_n$, corresponding to the same eigenvalue, as the ratio of two Wronskians
\begin{equation}
\hat\psi=\frac{\Wr[\psi_0,\psi_1, \dots, \psi_{n-1}, \psi]}{\Wr[\psi_0,\psi_1, \dots, \psi_{n-1}]} .
\label{dtNpsi}
\end{equation}
Meanwhile, the potential is transformed in the following way
\begin{equation}
u_n= u_0-2\left(\log\Wr[\psi_0,\psi_1, \dots, \psi_{n-1}]\right)''.
\label{dtNu}
\end{equation}
Here $\psi_i$ are $n$ eigenfunctions of the initial Schrodinger operator $H_0$ corresponding to the distinct eigenvalues $\eta_0,\eta_2,\dots,\eta_{n-1}$, i.e. $H_0\psi_i=\eta_i\psi_i$, $0\le i\le n-1$, called ``seed functions", and $\Wr$ stands for the Wronskian determinant. From (\ref{dtNpsi}), we get the Wronskian representation of the intertwining operator\footnote{Simplest proof of the Crum theorem uses the fact that, according to (\ref{DTpsi},\ref{HTTH}), $T_n$ annihilates $n$ seed functions $\psi_i$, $i=0,\dots, n-1$, i.e. kernel of $T_n$ is spanned by these seed functions. Then, since $T_n=\partial_z^n+\dots$, we immediately come to (\ref{TNW}).}
\begin{equation}
T_n[\,\boldsymbol{\cdot}\,]=\frac{\Wr[\psi_0,\psi_1, \dots, \psi_{n-1}, \,\boldsymbol{\cdot}\,]}{\Wr[\psi_0,\psi_1, \dots, \psi_{n-1}]} .
\label{TNW}
\end{equation}
In the particular case of confluent eigenvalues, when the spacing between $\eta_i$ vanishes, all $\eta_i$ tend to the same value $\eta$, and we have the chain of the type (\ref{rchain}), the Crum theorem degenerates to:
\begin{equation}
\varkappa_n=\frac{\Wr[\varkappa,\partial_\eta\varkappa, \dots, \partial_\eta^{n-1}\varkappa, \partial^n_\eta\varkappa]}{\Wr[\varkappa,\partial_\eta\varkappa, \dots, \partial_\eta^{n-1}\varkappa ]}, \quad u_n= u_0-2\left(\log\Wr[\varkappa,\partial_\eta\varkappa, \dots, \partial_\eta^{n-1}\varkappa]\right)'',
\label{varkappanu}
\end{equation}
where $\varkappa=\varkappa(z;\eta)$ is an eigenfunction of $H_0$ corresponding to the eigenvalue $\eta$, and $\varkappa_n$ is an eigenfunction of $H_n$ with the same eigenvalue. Here, the intertwining operator degenerates to :
$$
T_n[\,\boldsymbol{\cdot}\,]=\frac{\Wr[\varkappa,\partial_\eta\varkappa, \dots, \partial_\eta^{n-1}\varkappa, \,\boldsymbol{\cdot}\,]}{\Wr[\varkappa,\partial_\eta\varkappa, \dots, \partial_\eta^{n-1}\varkappa ]} .
$$
Eq. (\ref{varkappanu}) is useful for representing the Adler-Moser polynomials in the Wronskian form. Indeed, let us take derivatives of the Schrodinger equation $H_0\varkappa(z,\eta)=\eta\varkappa(z,\eta)$, $H_0=-\partial_z^2+u_0(z)$ wrt $\eta$ and then set $\eta=0$. In this way, we obtain the recurrence relation for derivatives $\partial^n_\eta \varkappa $ at the zero energy level:
$$
H_0\partial^n_\eta \varkappa\vert_{\eta=0}=n\partial^{n-1}_\eta \varkappa\vert_{\eta=0}, \quad H_0\varkappa\vert_{\eta=0}=0 .
$$
Thus, (\ref{varkappanu}) can be rewritten as
\begin{equation}
\varkappa_n=\frac{\tau_{n+1}}{\tau_n}, \quad \tau_n=\Wr[{\cal X}_1, {\cal X}_2, \dots, {\cal X}_{n-1}], \quad u_n=u_0-2(\log\tau_n)'' , \quad \tau_0=1 ,
\label{WXN}
\end{equation}
where the $\eta$-independent functions ${\cal X}_n(z)$ are defined by the recurrence relation
\begin{equation}
H_0{\cal X}_n=c_n{\cal X}_{n-1}, \quad H_0{\cal X}_1=0, \quad H_0=-\partial_z^2+u_0(z) ,
\label{H0Xn}
\end{equation}
In eq. (\ref{H0Xn}), $c_n$ are arbitrary constants. At each iteration of the Darboux transformation (\ref{muj}), apart from the common factor $C_i$, only one integration constant appear. Therefore, although, apart from $c_n$, two integration constant appear at each iteration of (\ref{H0Xn}), only one of them is esential\footnote{In other words, the sequence ${\cal X}_1$, ${\cal X}_2+c_2{\cal C}_1{\cal X}_1$, ${\cal X}_3+c_3{\cal C}_1{\cal X}_2+c_3c_2{\cal C}_2{\cal X}_1, \dots$, where ${\cal C}_i$ are arbitrary constants, is also a solution of (\ref{H0Xn}), and the values of Wronskians of this sequence $\Wr[{\cal X}_1, {\cal X}_2+c_2{\cal C}_1{\cal X}_1, \dots]$ do not depend on ${\cal C}_i$.}.

The intertwining operator is expressed in terms of ${\cal X}_i$ as
\begin{equation}
T_n[\,\boldsymbol{\cdot}\,]=\frac{\Wr[{\cal X}_1, {\cal X}_2, \dots, {\cal X}_{n-1}, \,\boldsymbol{\cdot}\,]}{\Wr[{\cal X}_1, {\cal X}_2, \dots, {\cal X}_{n-1}]} .
\label{TWronskian}
\end{equation}
When we start the factorization chain (\ref{rchain}) with the free Schrodinger operator $H_0=-\partial_z^2$ (i.e. with $u_0=0$), from (\ref{H0Xn}), we get
\begin{equation}
{\cal X}''_{n+1}(z)=(2n+1){\cal X}_n(z), \quad {\cal X}_1=z.
\label{Xn}
\end{equation}
Here\footnote{In (\ref{Xn}) we set $c_n=2n+1$. With this choice of normalization constants, $\tau_n$ is monic in $z$.}, the function ${\cal X}_n$ is a polynomial of degree $2n-1$ in $z$. Substituting $\varkappa_n=\tau_n/\tau_{n-1}$ (see(\ref{WXN})) into (\ref{muj}), we obtain recurrence relation for $\tau_n$ which coincides with that (\ref{AM3}) for the Adler-Moser polynomials $P_n$. Since the initial conditions of recurrence $\tau_0=P_0=1$, $\tau_1=P_1=z$ are the same for $\tau_n$ and $P_n$, we come to the conclusion that $\tau_n=P_n$. Thus, we obtain the Wronskian representation for the Adler-Moser polynomials
$$
P_n=\Wr[{\cal X}_1, {\cal X}_2, \dots, {\cal X}_n],
$$
where ${\cal X}_n$ are defined by (\ref{Xn}).

\section{Darboux Transformation and Equilibrium Configurations}
\label{dtequilibria}

According to (\ref{WXN}), in the case of Adler-Moser polynomials, iterations of Darboux transformations produce Schrodinger operators with potentials $u_n=-2(\log P_{n})''$ that are free of simple poles. Corresponding zero-level eigenfunctions $\varkappa_n=P_{n+1}/P_n$ are of the form $\prod_i (z-z_i)^{{\cal Q}_i}$, where ${\cal Q}_i=\pm 1$ are the values of charges in equilibrium, and $z_i$ are their positions.

Let us consider a more general situation, by introducing a factorizable eigenfunction
\begin{equation}
\psi(z)=\prod_{i=1}^N(z-z_i)^{{\cal Q}_i},
\label{psiQ}
\end{equation}
where, for the moment, we treat $z_i$ and ${\cal Q}_i$ as some unspecified numbers. Let us now look for all possible ${\cal Q}_i$ and $z_i$ for which $\psi$ satisfies the Schrodinger equation with the potential being free of simple poles:
$$
H\psi=(-\partial_z^2+u)\psi=0, \quad u \,\,\, \text{is free of simple poles} .
$$
Since
\begin{equation}
\frac{H\psi}{\psi}=u-(\log \psi)''-((\log \psi)')^2=u-\sum_i \frac{{\cal Q}_i}{(z-z_i)^2}-\sum_{i,j}\frac{{\cal Q}_i{\cal Q}_j}{(z-z_i)(z-z_j)},
\label{Hpsibypsi}
\end{equation}
the Schrodinger equation is satisfied if and only if the expression on the right-hand side of (\ref{Hpsibypsi}) vanishes identically in $z$. Applying the partial-fraction decomposition, we rewrite this expression as
\begin{equation}
u-\sum_i \frac{{\cal Q}_i({\cal Q}_i-1)}{(z-z_i)^2}-\sum_i \frac{2{\cal Q}_i\sum_{j\not=i}\frac{{\cal Q}_j}{z_i-z_j}}{z-z_i}.
\label{simplepolefree}
\end{equation}
For $u$ that is free of simple poles, (\ref{simplepolefree}) can vanish identically in $z$ only if all residues $-2{\cal Q}_i\sum_{j\not=i}\frac{{\cal Q}_j}{z_i-z_j}$ equal zero, i.e., only if $\sum_{j\not=i}\frac{{\cal Q}_j}{z_i-z_j}=0 $ for all $1\le i\le N$, which is precisely equilibrium conditions (\ref{static}). Hence, if $u$ is free of simple poles, ${\cal Q}_i$ and $z_i$ must satisfy (\ref{static}). The converse also follows from (\ref{simplepolefree}), so the equilibrium condition is equivalent to the absence of simple poles in the potential. From (\ref{simplepolefree}), we also obtain the potential
\begin{equation}
u=\sum_i \frac{{\cal Q}_i({\cal Q}_i-1)}{(z-z_i)^2}.
\label{u}
\end{equation}

Now, suppose that there exists a factorizable Darboux transform $\hat \psi$ of $\psi$:
\begin{equation}
\hat\psi(z)=\prod_{i=1}^{\hat N}(z-\hat z_i)^{\hat{\cal Q}_i} .
\label{tildepsi}
\end{equation}
Then $\hat\psi$ also corresponds to a static configuration.

Here $\hat N$ charges $\hat{\cal  Q}_1, \dots, \hat{\cal Q}_N$ are in equilibrium at points $\hat z_1, \dots, \hat z_{\hat N}$. Indeed, equilibrium conditions for $\hat {\cal Q}_i$, $\hat z_i$ are satisfied if the potential $\hat u$ of the Darboux transformed Schrodinger operator $\hat H$ is free of simple poles. According to (\ref{DTH}), transform of $u$ equals
\begin{equation}
\hat u=u-2(\log \psi)'' .
\label{dtupsi}
\end{equation}
Then, from (\ref{psiQ}) and (\ref{u}) we obtain
\begin{equation}
\hat u=\sum_i \frac{{\cal Q}_i({\cal Q}_i+1)}{(z-z_i)^2},
\label{hatu}
\end{equation}
i.e. $\hat u$ is a sum of second order poles. Therefore, $\hat\psi$, of the form (\ref{tildepsi}), corresponds to a static configuration.

Note, that since 
$$
\hat u=\sum_i \frac{{\cal \hat Q}_i({\cal \hat Q}_i-1)}{(z-\hat z_i)^2},
$$
from (\ref{hatu}), we see that the Darboux transformation can act on charges in two ways
\begin{equation}
\hat{\cal Q}= -{\cal Q}, \quad \hat{\cal Q} = {\cal Q}+1, 
\label{QQhat}
\end{equation}
i.e., it either (i) inverts a charge or (ii) increments a charge by 1. In particular, it can create new charges out of ``zero charges" $\hat{\cal Q}=0+1=1$ and annihilate charges with ${\cal Q}=-1$, $\hat{\cal Q}=-1+1=0$. In the case of the Adler-Moser polynomials in generic configurations, it annihilates negative charges, inverts positive charges, and creates positive charges at new positions.

We also note that, when all charges belong to the set
$$
{\cal Q}_i \in \left\{ -1, \frac{1}{2}, 1, \frac{3}{2}, 2, \dots\right\},
$$
the zero-level transform $\hat\psi=\frac{C}{\psi}\int\psi^2 dz$ of $\psi$ is factorizable \cite{KWCC,ON2}: Here, $\psi^2$ is rational and has poles only at $z=z_i$, corresponding to ${\cal Q}_i=-1$. Estimating residues of $\psi^2$ at these positions, one concludes that they equal zero due to (\ref{static}). Therefore, $\int\psi^2dz$ is rational, and $\hat\psi$ corresponds to an equilibrium configuration when all ${\cal Q}_i$ belong to the above set.

The presence of charges ${\cal Q}_i=-1/2$ always leads to logarithmic terms in $\hat\psi$.

\section{Configurations Related to Even Bispectral Family}
\label{Bispectral}

In this section, we review a family of factorizable eigenfunctions generated through Darboux transformations, whose generic members correspond to the equilibria of three species of charges \cite{DG,KWCC,ONCS}.

We recall that the sequence $H_i$ of Schrodinger operators, corresponding to the Adler-Moser polynomials begins with the free Schrodinger operator $H_0=-\partial_z^2$. This sequence is known to constitute the ``odd" family of bi-spectral operators\footnote{The differential operator $H$
 in $z$ is bi-spectral if there exists a differential operator $B$ in $k$ and a common eigenfunction, such that  $H\psi(z,k)=f(k)\psi(z,k)$, $B\psi(z,k)=g(z)\psi(z,k).$ For the even family of Schrodinger operators, the common eigenspace of $H$ and $B$ is two-dimensional, while in the odd case, it is generally one-dimensional.} \cite{DG}. There also exists another sequence of the Darboux transforms, which forms the ``even" family of bi-spectral operators \cite{DG}. This sequence starts with the operator
\begin{equation}
H_0=-\partial_z^2-\frac{1}{4z^2}.
\label{H0even}
\end{equation}
In both the ``even" and ``odd" cases we present elements of the sequence of zero-level eigenfunctions of the Darboux-transforms $H_i$:
$$
\psi_0\to\psi_1\to\psi_2\to\dots, \quad H_i\psi_i=0
$$
as the ratios (cf. (\ref{WXN}))
\begin{equation}
\psi_i=P_{i+1}/P_i , \quad P_0=1 .
\label{hi}
\end{equation}
According to section \ref{DTFC} (see eq. (\ref{DT0})), the Darboux transform $\psi_{i+1}$ of $\psi_i$ is
\begin{equation}
\psi_{i+1}=\frac{C_i}{\psi_i}\int \psi_i^2 dz.
\label{echain}
\end{equation}
From this and (\ref{hi}), we derive the differential recurrent relation for $P_i$:
\begin{equation}
P'_{i+1}P_{i-1}-P'_{i-1}P_{i+1}=C_iP^2_i .
\label{Pi}
\end{equation}

In the ``odd" (Adler-Moser) case $P_0=1$, $P_1=z$ (see eq. (\ref{AM3})). In the ``even" case $H_0[z^{1/2}]=0$ (i.e. $\psi_0=z^{1/2}$ see (\ref{H0even})) and, according to (\ref{hi})
\begin{equation}
P_0=1, \quad P_1=z^{1/2} .
\label{P0P1even}
\end{equation}
Note that, for convenience, we set the normalization constants $C_i$ in (\ref{Pi}) to values that make $P_i(z)$ monic\footnote{In the even case, this differs from the original normalization of \cite{DG}, where same $C_i=2i+1$ were chosen for both families.}. For the ``even" family, from (\ref{Pi},\ref{P0P1even}) we get $P_2=z^2+s_1$. Thus, using (\ref{hi}), we obtain the sequence of length 2 depending on one parameter $s_1$:
\begin{equation}
\psi_0=z^{1/2}, \quad \psi_1=\frac{z^2+s_1}{z^{1/2}} .
\label{varkappa01}
\end{equation}
Another application of the recurrence relation (\ref{Pi}) with $s_1\not=0$ would produce a logarithmic term. Therefore, to obtain $\psi_2$ and $\psi_3$ of the form (\ref{psiQ}), we restrict $s_1$ to be zero and pick a new integration constant $s_2$:
\begin{equation}
P_3=z^{9/2}+s_2z^{1/2}, \quad
P_4=z^8+6s_2z^4+s_3z^2-3s_2^2.
\label{P3P4}
\end{equation}
Thus, we obtain the sequence of length 4:
$$
\psi_0=z^{1/2}, \quad \psi_1=z^{3/2}, \quad \psi_2=\frac{z^4+s_2}{z^{3/2}}, \quad \psi_3=\frac{z^8+6s_2z^4+s_3z^2-3s_2^2}{z^{9/2}+s_2z^{1/2}}
$$
depending on two parameters, $s_2$ and $s_3$.

Next, to eliminate logarithmic terms in $P_5$ and $P_6$, one has to set both $s_1=s_2=0$. We then obtain a sequence $\psi_0, \psi_1, \psi_2, \psi_3, \psi_4, \psi_5$ (of length 6) depending on three parameters $s_3$, $s_4$, and $s_5$. And so forth.

Sequences of the even family have finite lengths (i.e., they terminate) due to the presence of charge of the third species at the origin: Initially, charge $1/2$ is placed at $z=0$. According to (\ref{QQhat}), the following transformations occur with this charge: It is incremented by $1$ during each of the $m-1$ first Darboux transformations. It is inverted by the $m$th transformation and becomes equal to $-m+1/2$. The charge is then incremented again during the next $m-1$ steps, until it reaches value $-1/2$. This charge would create a logarithmic term in (\ref{echain}) at the next step, so the sequence terminates.

Closing this section, we note that similarly to the odd family, the Wronskian representation of $\psi_i$ can also be obtained for the even family by applying the procedure (\ref{WXN},\ref{H0Xn}). This will be presented in Section \ref{streetseven} (eqs. (\ref{tauiX}) and (\ref{XNeven})).

\section{KWCC Transformation and Terminating Configurations}
\label{KWCCterminating}

Let us now return to the sequence of transformations of rational functions (whose primitives are also rational functions) corresponding to \footnote{Or to the chain of solutions $\dots\to p_{i-1}, q_i \to  p_i,q_i \to p_i,q_{i+1}\to p_{i+1},q_{i+1} \to \dots$ of (\ref{bilinear}).} iterations (\ref{qup},\ref{pup})
\begin{equation}
\dots\to \frac{q_i^{2\Lambda}}{p_{i-1}^2} \to  \frac{p_i^{2/\Lambda}}{q_i^2} \to \frac{q_{i+1}^{2\Lambda}}{p_i^2}\to\frac{p_{i+1}^{2/\Lambda}}{q_{i+1}^2} \to \dots.
\label{chain}
\end{equation}
Denoting the function $\frac{q_i^{2\Lambda}}{p_{i-1}^2}$ from the above chain as $\psi^2$ and the next function $\frac{p_i^{2/\Lambda}}{q_i^2}$ as $\tilde\psi^2$,  from (\ref{pup}) we get the relationships between them
\begin{equation}
\psi\to\tilde\psi, \quad \tilde\psi(z)=C \left(\frac{1}{\psi(z)}\int \psi(z)^2dz\right)^\gamma,
\label{KWCC}
\end{equation}
where, for this step of the chain, $\gamma=1/\Lambda$ and $C$ stands for a non-zero constant.

The next transformation in the sequence  (\ref{chain}) relates $\psi^2=\frac{p_i^{2/\Lambda}}{q_i^2}$ and $\tilde\psi^2=\frac{q_{i+1}^{2\Lambda}}{p_i^2}$. Using (\ref{qup}), we also arrive to transformation (\ref{KWCC}), this time with $\gamma=\Lambda$.

Therefore, the chain (\ref{chain}) can be constructed by the iterative applications of the following transformations
\begin{equation}
\psi_{i+1}=C_i\left(\frac{1}{\psi_i(z)}\int \psi_i(z)^2dz\right)^{\gamma_{i+1}} ,
\label{KWCCi}
\end{equation}
where
\begin{equation}
\gamma_{i+1}=1/\gamma_i,
\label{gammai}
\end{equation}
and
\begin{equation}
\gamma_i \in \{1/2, 1, 2\} .
\label{gi}
\end{equation}
The transformation (\ref{KWCC}) was presented by Krishnamurthy, Wheeler, Crowdy and Constantin (KWCC) in \cite{KWCC}.

The KWCC transformation (\ref{KWCC}) is a composition of two mappings

\begin{enumerate}[(i)]
 \item The Darboux transformation at zero energy level $ \hat\psi=\frac{C}{\psi}\int \psi^2dz $ (see (\ref{DT0}) ) and
 \item Exponentiation\footnote{When $\gamma=-1$, the exponentiation $\psi\to 1/\psi$ is a special case of the zero-level Darboux transformation. In this case, we have the composition of two Darboux transformations. When we compose the transformations in the reverse order, the result is $\psi\to C\psi\int\psi^{-2}dz$, which is a transformation between two linearly independent solutions of the same Schrodinger equation. From this, it follows that iterations of the KWCC transformations with $\gamma=-1$ merely reparametrize the result of the first transformation.}  $\tilde\psi=\hat\psi^\gamma$.
\end{enumerate}

According to section \ref{dtequilibria}, the Darboux transformation is a mapping between static configurations, provided the reuslt of transformation, $\hat \psi$, is factorizable, i.e., it has a form (\ref{tildepsi}). The exponentiation corresponds to scaling of all charges $\hat {\cal Q}_i \to \gamma\hat {\cal Q}_i$, and is also a mapping between static configurations. Therefore, when $\int\psi^2 dz$ is rational, the KWCC transformation is also such a mapping. Rationality of primitive of $\psi_i^2$ imposes restrictions on $\gamma_i$ and on constants of integration.

From the above and (\ref{QQhat}), it follows that KWCC transformation (\ref{KWCC}) can act on charges in two ways
\begin{equation}
\tilde {\cal Q}= -\gamma {\cal Q}, \quad \tilde {\cal Q} = \gamma\left({\cal Q}+1\right) ,
\label{QQgamma}
\end{equation}
which includes the creation of new charges out of ``zero charges" at new positions, where $\tilde{\cal Q}=\gamma(0+1)=\gamma$. Obviously, the KWCC transformations (\ref{KWCCi},\ref{gammai},\ref{gi}) generate all classes of configurations considered above, i.e. those related to the odd and even bi-spectral families as well as to the $\Lambda=2$ case.

To rewrite the sequence of transforms (\ref{KWCCi}), (\ref{gammai}) in terms of polynomials, we present $\psi_i$ as
$$
\psi_i=\frac{\tau_i^{\gamma_i}}{\tau_{i-1}},
$$
where $\tau_i$ is a polynomial of an integer or half-integer degree $d_i$. Here, a ``polynomial of half-integer degree" in $z$ means a polynomial in $z$ times $z^{1/2}$. Then from (\ref{KWCCi}) and (\ref{gammai}) we get
$$
\tau_{i+1}=(2\gamma_id_i-2d_{i-1}+1)\tau_{i-1}\int\frac{\tau_i^{2\gamma_i}}{\tau_{i-1}^2}dz,
$$
where
\begin{equation}
d_{i+1}=2\gamma_id_i-d_{i-1}+1, \quad \gamma_{i+1}=1/\gamma_i, \quad d_0=0.
\label{degrees}
\end{equation}
In the differential form these relations write as
\begin{equation}
\tau_{i+1}'\tau_{i-1}-\tau_{i+1}\tau_{i-1}'=(2\gamma_id_i-2d_{i-1}+1)\tau_i^{2\gamma_i}, \quad \tau_i(z)=z^{d_i}+\dots , \quad \tau_0=1 .
\label{Bartmantau}
\end{equation}
These are relations (\ref{Bartman}), but now $d_i$ are not necessarily the degrees (\ref{lmlambda1}) or (\ref{lmlambda2}) of the main sequences. This is because (\ref{degrees}, \ref{Bartmantau}) are derived from the KWCC chain, rather than from the bi-linear equation (\ref{bilinear}). Because of this, any polynomial solution of (\ref{Bartmantau}) that starts from a static configuration (e.g. from a single vortex) also corresponds to a sequence of static configurations.

Apart from the infinite main sequences of the $\Lambda=1$ and $\Lambda=2$ cases, recurrence relations (\ref{degrees},\ref{Bartmantau}) allow one to find sequences of configurations of finite lengths, i.e., terminating configurations. The ``even" bi-spectral family of section \ref{Bispectral}, i.e., solutions of (\ref{degrees},\ref{Bartmantau}) corresponding to $\gamma_i=1$, $d_1=1/2$, is a family of terminating configurations for $\Lambda=1$. In this case, (\ref{Bartmantau}) are relations (\ref{Pi}). 

In the $\Lambda=2$ case, Eqs. (\ref{Bartmantau}) are, modulo normalization factors, relations (\ref{qqp_ppq}). As an example of a family of terminating configurations for $\Lambda=2$, we can take the case $\gamma_0=2$, $d_1=2$, i.e., $\tau_1=z^2$. Here are several first terminating sequences of $\tau_i$:
$$
\begin{array}{l}
1, z^2, z^3+s_1 \\
1, z^2, z^3, z^2(z^9+r_1), z^9+s_2 z^3-2r_1 \\
1, z^2, z^3, z^{11}, z^3(z^6+s_2), z^2(z^{24}+\frac{20}{3}z^{18}s_2+30z^{12}s_2^2+z^9r_2-20z^6s_2^3-\frac{5}{3}s_2^4), \\
\qquad\qquad\qquad\qquad\qquad\qquad\qquad\qquad\qquad\qquad z^{18}+15z^{12}s_2+z^9s_3-45z^6s_2^2+\frac{2s_2s_3-3r_2}{2}z^3+5s_2^3 \\
\dots .
\end{array}
$$
Similarly to the even bi-spectral family, here, a charge of the third species undergoes transformations (\ref{QQgamma}) at $z=0$ (this time with $\gamma\in\{1/2,2\}$) until the logarithmic singularity is encountered in (\ref{KWCCi}).

One can compute terminating sequences by solving (\ref{Bartmantau}) for $\tau_{i+1}$ as for a monic polynomial with unknown coefficients. Compatibility of resulting system of algebraic equations (which is linear in coefficients of $\tau_{i+1}$) imposes restrictions on the free parameters of $\tau_i$. For sequences related to the even bi-spectral family, there exists a Wronskian representation (see Section \ref{streetseven}).

 We recall that the main sequences exhaust all possible polynomial solutions of (\ref{bilinear}) with no common/multiple roots for the  cases $\Lambda=1$ and $\Lambda=2$. Solutions of (\ref{Bartmantau}) that do not belong to the main sequences do not satisfy the bi-linear equation (\ref{bilinear}); instead, they satisfy its generalization.
 
 In more detail: Let us first consider a sequence of the Darboux transforms at the zero energy level, starting from $H_0=-\partial_z^2+u_0$, with non-zero $u_0\not=0$. At the $n$th step of the chain, we have
 $$
 u_n=u_0-2(\log\tau_n)'', \quad \varkappa_n=\frac{\tau_{n+1}}{\tau_n}, \quad \tau_0=1, \quad H_n\varkappa_n=0 .
 $$
 Therefore
 $$
 H_n\varkappa_n=\left(-\partial_z^2+u_0-2(\log\tau_n)''\right)\left[\frac{\tau_{n+1}}{\tau_n}\right]=0 ,
 $$
 or equivalently
 \begin{equation}
 \tau_{n+1}''\tau_n-2\tau_{n+1}'\tau_n'+\tau_{n+1}\tau_n''-u_0\tau_n\tau_{n+1}=0.
 \label{rchainu0}
 \end{equation}
 This is a generalization of the Tkachenko equation (\ref{Tkachenko}) that includes cases of non-integer $d_i$. For the even bi-spectral family, the initial potential equals to $u_0=a(a-1)/z^2$, where $a$ is a half-integer. Then, substituting $\tau_n=p$, $\tau_{n+1}=z^a q$ into (\ref{rchainu0}), we get
 $$
 p''q-2p'q'+pq''+\frac{2a}{z}(q'p-p'q)=0 .
 $$
 In fact, this generalization is a particular case of the tri-linear equation for three species of vortices with ${\cal Q}_i \in \{-1,1,a\}$, where a single vortex of the third species of strength $a$ is placed at the origin $z=0$ \cite{DK1,ONCS}. In general, for multiple species we have the poly-linear equation (see e.g. \cite{Lbhe,ON2})
 \begin{equation}
 \sum_i\Lambda_i^2\frac{{\cal P}_i''}{{\cal P}_i}+2\sum_{i<j}\Lambda_i\Lambda_j\frac{{\cal P}_i'}{{\cal P}_i}\frac{{\cal P}_j'}{{\cal P}_j}=0 ,
 \label{multilinear}
 \end{equation}
 where sums run over all species, $\Lambda_i$ is a strength of the $i$th species, and ${\cal P}_i$ is polynomial whose roots correspond to the coordinates of vortices of this species. In the case of configurations related to the even bi-spectral family ${\cal P}_1=p$, ${\cal P}_2=q$, and ${\cal P}_3=z$. In a similar situation, when ${\cal Q}_i \in \{-1, \Lambda, \Lambda a\}$, the following generalization of Eq. (\ref{bilinear}) holds
 $$
 p''q-2\Lambda p'q'+\Lambda^2 pq''+\frac{2a\Lambda}{z}(\Lambda q'p-p'q)=0 .
 $$
 Exploring this equation, O'Neil and Cox-Steib \cite{ONCS} presented the families of terminating configurations.

\section{$\Lambda=2$ Case and Third-Order Operators}
\label{3rd_order}

In the $\Lambda = 1$ case, the KWCC transformation (\ref{KWCCi}) is a Darboux transformation at the zero energy level (\ref{echain}), since here $\gamma_i = 1$. However, when $\Lambda = 2$, i.e., $\gamma_i \in \{1/2, 2\}$, the KWCC transformation is a composition of the Darboux transformation and exponentiation. This raises a natural question about the existence of a Darboux transformation in the $\Lambda = 2$ case. The answer to this question is affirmative. However, unlike the $\Lambda = 1$ case, one must consider Darboux transformations of third-order differential operators rather than second-order Schrödinger operators \cite{LY1}.

In more detail: The $\Lambda=2$ specification of the bilinear equation (\ref{bilinear}) can be written in the Schrodinger form as
$$
\left(-\partial_z^2-6(\log q)''\right)\left[\frac{p}{q^2}\right]=0.
$$
As follows from Section \ref{L12}, for $q=q_n$, two linearly independent solutions of the above equation are $p_{n-1}/q_n^2$ and $p_n/q_n^2$ respectively. Since a linear combination of $p_n$ and $p_{n-1}$ is just a reparametrized $p_n$ or $p_{n-1}$, we can write 
\begin{equation}
\left(-\partial_z^2+u_n\right)\phi=0, \quad u_n=-6(\log q_n)'' ,
\label{Schrodinger2}
\end{equation}
where
\begin{equation}
\phi=p_n/q_n^2 \quad {\rm or} \quad \phi=p_{n-1}/q_n^2.
\label{varkappa2}
\end{equation}
In contrast to the $\Lambda=1$ case, the above sequence of second-order operators cannot be generated by Darboux transformations (\ref{DTH}), because now $u_{n\pm 1}\not=u_n-2(\log\phi)''$. To proceed, we note that according to the first-order recurrence relations (\ref{qqp_ppq}), $p_n$ in (\ref{Schrodinger2}, \ref{varkappa2}) can be replaced with $q_{n+1}'q_n-q_{n+1}q_n'$, so $\phi=(q_{n\pm 1}/q_n)'$. Then, from (\ref{Schrodinger2}) it follows that the ratio  $q_{n\pm 1}/q_n$ satisfies the third-order differential equation
$$
L\left[\frac{q_{n\pm 1}}{q_n}\right]=0, \quad L=\partial_z^3-u_n\partial_z .
$$
The general solution $\varkappa$ of this differential equation, i.e., the general solution of
\begin{equation}
L\varkappa=0, \quad L=\partial_z^3-u_n\partial_z ,
\label{Ln}
\end{equation}
is a linear combination of three functions: $\left\{q_{n+1}/q_n, q_{n-1}/q_n, 1\right\}$ . These functions are linearly independent because $q_{n-1}, q_n, q_{n+1}$ have distinct degrees.

Any non-constant solution of (\ref{Ln}), i.e., $\varkappa=q/q_n$, where $q=C_1q_{n+1}+C_0q_n+C_{-1}q_{n-1}$, corresponds to an equilibrium configuration, since $\phi=\varkappa'=p/q_n$, with $p=q'q_n-q_n'q$ being a polynomial. The polynomial $q=C_1q_{n+1}+C_0q_n+C_{-1}q_{n-1}$ is a reparametrized $q_{n\pm 1}$ or reparametrized $q_n$.

Thus, the set $\{C_n^\pm q_{n\pm 1}/q_n$: $n\in\mathbb{Z}\}$, where $C_n^\pm $ are arbitrary constants, contains all solutions\footnote{The constant solution corresponds to the limit $C_n^\pm\to 0$, $s_{\pm n}\to\infty$, $C_{\pm n}^\pm s_{\pm n}={\rm const}$, $n>0$ (see (\ref{sr})).} of (\ref{Ln}). It corresponds to a complete set of equilibrium configurations of two species of charges with ${\cal Q}_i\in\{-1,2\}$. The zeros of $\phi=\varkappa'=(q_{n \pm 1}/q_n)'$ correspond to the positions of charges with ${\cal Q}_i=-1$, while its poles correspond to positions of charges with ${\cal Q}_i=2$.

\section{Darboux Transformation for Third-Order Operators and Equilibrium Configurations}
\label{3rd_order_equilibrium}

Darboux transformations for third-order operators of the form $\partial_z^3-u\partial_z$ were found by Aiyer et al. \cite{ABO}. Their derivation, via factorization, was presented by Athorne and Nimmo in \cite{AtNi}. Here, we will derive a zero-level chain of Darboux transformations for such operators. 

We consider the third-order operator $L$ and its transform $\hat L$
\begin{equation}
L=\partial_z^3-u\partial_z, \quad \hat L=\partial_z^3-\hat u\partial_z .
\label{DT3}
\end{equation}
Let $\varkappa$ be a non-constant element of kernel of $L$. Operator $L$ can be expressed as the product of the second and the first-order factors
\begin{equation}
L=BA,
\label{LBA}
\end{equation}
where
\begin{equation}
B=\partial_z^2+v\partial_z-v'-\frac{v''}{v}, \quad A=\partial_z-v, \quad v=\frac{\varkappa'}{\varkappa} .
\label{BA}
\end{equation}
The potential $u$ is expressed in terms of $v$ as follows
\begin{equation}
u=3v'+v^2+\frac{v''}{v} .
\label{u3}
\end{equation}
In contrast to the case of the Schrodinger operator, permutation of factors in (\ref{LBA}) maps $L$ into $\hat L$ of the similar type -- that is, an operator without the zeroth-order term, as in (\ref{DT3}) -- only if\footnote{This corresponds to $u={a}^{2} \left( 6\,c\,{\rm cn} \left( az+b,c \right) {\rm dn}\left( az+b,c \right)+6\,{c}^{2} \, {\rm sn}^2 \left( az+b,c \right) -{c}^{2} - 1 \right)$ and $\hat u=a^2( 6\,{c}^{2} {\rm sn}^2 \left( az+b,c\right)-{c}^{2}-1)$. In particular, in the rational limit, $u=0$ or $u=12/z^2$, where $v=2/z$ or $v=-2/z$ respectively.} $v=2ac\,{\rm sn}(az+b,c)$, where $a,b$ and $c$ are arbitrary constants. In general, to obtain an operator of the same type, several permutations and re-factorizations are required (see Appendix \ref{AA} for more details). Instead of studying these intermediate operations separately, we consider their result:
\begin{equation}
\hat L=\hat B \hat A ,
\label{hatLBA}
\end{equation}
where $\hat B$ and $\hat A$ are of the form (\ref{BA}) with $v$ replaced by some $\hat v$:
\begin{equation}
\hat B=\partial_z^2+\hat v\partial_z-\hat v'-\frac{\hat v''}{\hat v}, \quad \hat A=\partial_z-\hat v .
\label{hatBA}
\end{equation}
We recall that in the Schrodinger operator case  (cf. (\ref{AstarA})), the Darboux transformation was the involution $A=\partial_z-v \leftrightarrow A^*=-(\partial_z+v)$, which corresponds to permutation of factors in the Schrodinger operator. This is, in fact, the involution $v\to -v$. Let us apply similar involution:
$$
\hat v = -v
$$
to our third-order operators. Then, from (\ref{BA}, \ref{u3}), it follows that for such a transformation
\begin{equation}
\hat u = u - 6 (\log \varkappa)'' .
\label{DTu3}
\end{equation}
Since $\hat A=\partial+v=\partial+\varkappa'/\varkappa$, the function $\varkappa^{-1}$ is an eigenfunction of $\hat L=\hat B\hat A$ corresponding to the zero eigenvalue:
$$
\hat L \varkappa^{-1}=0 .
$$
The complete zero-level transform $\hat\varkappa$ of $\varkappa$ is the general solution of equation
\begin{equation}
\hat L\hat\varkappa=0.
\label{hLhkappa}
\end{equation}
It can be obtained from its particular solution $\varkappa^{-1}$ by elementary methods: Fist, we rewrite (\ref{hLhkappa}) as
$$
\hat \phi''-\hat u \hat \phi=0, \quad \hat\phi=\hat\varkappa' .
$$
One of the linearly independent solution of the above second order equation equals $(\varkappa^{-1})'$. Then, its general solution is
\begin{equation}
\hat\phi = C(\varkappa^{-1})'\int \frac{dz}{\left((\varkappa^{-1})'\right)^2} ,
\label{hatphi}
\end{equation}
where $C$ is an arbitrary constant and the primitive includes an arbitrary constant of integration. To obtain $\hat\varkappa$, we have to integrate (\ref{hatphi}), i.e.
$\hat\varkappa=\int\hat\phi dz=C\int \left(\int \frac{dz}{\left((\varkappa^{-1})'\right)^2}\right)d\frac{1}{\varkappa}$. Integrating by parts, we obtain, modulo multiplication by an arbitrary constant,
\begin{equation}
\hat\varkappa = \int \frac{\varkappa^3}{(\varkappa')^2}dz - \frac{1}{\varkappa}\int \frac{\varkappa^4}{(\varkappa')^2}dz,
\label{DTkappa}
\end{equation}
where the primitives include arbitrary constants of integration. Thus, we have found the zero-level Darboux transformations (\ref{DTu3}, \ref{DTkappa}) for the third-order operators of the type $\partial_z^3-u\partial_z$.

In the case where $\varkappa$ corresponds to an equilibrium configuration and its transform $\hat\varkappa$ is factorizable, $\hat\varkappa$ also corresponds to an equilibrium configuration. This can be shown by arguments that are slight modifications of those in Section \ref{dtequilibria}. Specifically, we need to replace $\psi$, $\hat\psi$ from Section \ref{dtequilibria} with\footnote{Obviously, derivative of factorizable function is also factorizable.} $\phi=\varkappa'=\prod_i (z-z_i)^{{\cal Q}_i}$, $\hat\phi=\hat\varkappa'=\prod_i (z-\hat z_i)^{\hat{\cal Q}_i}$, and replace the Darboux transformation for potential (\ref{dtupsi}) with the transformation (\ref{DTu3}).

Up to multiplications by normalization constants, the transformations (\ref{DTkappa}) are equivalent to the first-order recurrence relations (\ref{qqp_ppq}). Therefore (as discussed in section \ref{KWCCterminating}) they generate main\footnote{Actually, for the main sequences this immediately follows from (\ref{DTu3}) and the fact that $\varkappa_n=q_{n\pm 1}/q_n$ is the general solution of (\ref{Ln}) with $u_n=-6(\log q_n)''$.} as well as terminating sequences in the $\Lambda=2$ case\footnote{A similar picture holds in the $\Lambda=1$ case, where the zero-level Darboux transformation $\hat\psi=C\psi^{-1}\int\psi^2dz$, $\hat\psi=P_+/P $, $\psi=P/P_-$ is equivalent to the recurrence relation $P_+'P_--P_+P_-'=CP^2$.}. Indeed, presenting $\varkappa$ and $\hat\varkappa$ as
\begin{equation}
\hat\varkappa=q_+/q, \quad \varkappa=q/q_- ,
\label{qpmq}
\end{equation}
where $q_+$, $q$ and $q_-$ are some functions of $z$, we write
\begin{equation}
\hat\varkappa'=p/q^2, \quad \varkappa'=p_-/q_-^2 ,
\label{kprime}
\end{equation}
where
\begin{equation}
p = q_+'  q-q_+ q', \quad p_- = q'  q_- - q q_-' .
\label{Abel1}
\end{equation}
Substituting (\ref{qpmq},\ref{kprime}) into transformation (\ref{DTkappa}), or equivalently, substituting $\hat \phi=p/q^2$, $(\varkappa^{-1})'=-\varkappa^{-2} p_-/q_-^2$ and $\varkappa=q/q_-$ into (\ref{hatphi}) and rewriting result in the differential form, we obtain
\begin{equation}
q^4 \propto p'  p_- - p p_-' .
\label{Abel2}
\end{equation}
Equations (\ref{Abel1}, \ref{Abel2}) are, up to normalization factors, the first-order recurrence relations (\ref{qqp_ppq}) with $q_+=q_{n+1}$, $q=q_n$, $q_-=q_{n-1}$ and $p=p_n$, $p_-=p_{n-1}$.

Therefore, the iterations $\varkappa_{n+1}=\hat\varkappa_n$ of the Darboux transformations (\ref{DTkappa}) generate main, as well as terminating sequences in the $\Lambda=2$ case, where equilibrium positions of charges are zeros and singular points of $\varkappa_n'$.

Closing this section, we note that the operators $L_n=(\partial_z^2-u_n)\partial_z=\partial_z^3+6(\log q_n)''\partial_z$ (see (\ref{Schrodinger2},\ref{Ln})), which correspond to the main $\Lambda=2$ sequences,
are rational Lax operators of the Sawada-Kotera hierarchy. The functions $q_n$ are polynomial $\tau$-functions of the hierarchy. Such $\tau$-functions can be written in a Pfaffian form \cite{KVdL}. Details are provided in Appendix \ref{AA}, where we also present intertwining operators for $L_n$. Looking ahead, we note that, in contrast to $\Lambda=1$ case, these intertwining operators cannot be used to obtain translating configurations. This is because they intertwine third-order operators, rather than second-order Schrodinger operators. We will discuss the implications of this fact in the conclusion section.

In the next section we will focus on translating configurations in the $\Lambda=1$ case.

\section{Translating Configurations and Baker-Akhieser Functions}
\label{translating}

Translating configurations of vortices are solutions of (\ref{stationary}). To generate them, we consider the following eigenfunction
\begin{equation}
\psi(k,z)=C(k)\prod_{i=1}^N(z-z_i(k))^{{\cal Q}_i}e^{kz},
\label{psik}
\end{equation}
of the Schrodinger operator $H$:
$$
H\psi=(-\partial_z^2+u)\psi=-k^2\psi .
$$
The condition that the potential $u=\psi''/\psi-k^2$ does not have simple poles (i.e., it is of form (\ref{u})) is equivalent to the equilibrium conditions (\ref{stationary}) for a translating configuration. This can be shown by arguments similar to those employed for the static configurations in Section \ref{dtequilibria}.

In the case of the factorization chain (\ref{rchain}) starting with the free Schrodinger operator, there exists an eigenfunction, called the ``rational Baker-Akhieser function"\footnote{In the theory of integrable hierarchies of non-linear PDEs, the Baker-Akhieser function is a solution of an infinite system of linear PDE's. Associated integrable hierarchy arises as a condition of compatibility of the system. Here we deal with the Baker-Akhieser function of the KdV hierarchy.}, which has the form (\ref{psik}). It is an eigenfunction of $H_j$, which is obtained by the action of an intertwining operator with rational coefficients on the eigenfunction  $e^{kz}$ of the free Schrodinger operator, i.e., $\Psi=C(k)T_j[e^{kz}]$. Since the $z$-independent factor $C(k)$ does not play role in our considerations, we will use the eigenfunction
\begin{equation}
\psi_j=T_j[e^{kz}]=\frac{p_j}{q_j}e^{kz} ,
\label{Te}
\end{equation}
where $p_j$ and $q_j$ are polynomials in $z$. As we have shown in section \ref{Bispectral}, the corresponding potential $u_j$, obtained by $j$ Darboux transformations from $u_0=0$, is of the  form (\ref{u}). Since $u_j$ does not contain simple poles, eigenfunction (\ref{Te}) corresponds to a translating configuration. It will be shown in section \ref{sources}  that the class of such configurations includes all possible translating patterns consisting of finite number of vortices with ${\cal Q}_i=\pm 1$.

A constructive way to write down $\psi_j$ is to use the Wronskian representation of $T_j$, given by (\ref{TWronskian}, \ref{Xn}). Then, from (\ref{Te}), we get
$$
\psi_j=\frac{\Wr \left[{\cal X}_1,{\cal X}_2,\dots,{\cal X}_j,e^{kz}\right]}{\Wr \left[{\cal X}_1,{\cal X}_2,\dots,{\cal X}_j\right]}, \quad {\cal X}_1=z, \quad {\cal X}''_j=(2j+1){\cal X}_{j-1} ,
$$
and from (\ref{Te}) it follows that
\begin{equation}
p_j= e^{-kz}\Wr \left[{\cal X}_1,{\cal X}_2,\dots,{\cal X}_j,e^{kz}\right], \quad q_j=\Wr \left[{\cal X}_1,{\cal X}_2,\dots,{\cal X}_j\right]= P_j,
\label{AMtranslating}
\end{equation}
where $P_j$ is the $j$th Adler-Moser polynomial. When $k\not=0$, the degree of $q_j(z)$ equals to that of $p_j(z)$. This corresponds to the neutrality condition (\ref{neutrality}) for an equilibrium in the homogeneous electric field $k\not=0$. In the $k\to 0$ limit
$$
\psi_j(k=0,z)={\rm const}\frac{P_{j-1}}{P_j},
$$
and we get the static equilibrium with non-zero total charge. In other words, in the $k\to 0$ limit a part of positive charges moves to infinity.

\section{Equilibria in Non-Homogenous Background Fields}
\label{nonuniform}

Let us assume that the Schrodinger operator $H=-\partial_z^2+u(z)$ has the following ``quasi-rational" eigendunction
\begin{equation}
\psi(z)=\prod_{i=1}^N(z-z_i)^{{\cal Q}_i} e^{\Phi(z)}, \quad H\psi=\lambda\psi,
\label{psi2}
\end{equation}
where $\Phi(z)$ is a polynomial. ``Quasi-rational" here means that all $Q_i$ are integers. Requirement that potential $u=\frac{\psi''}{\psi}+\lambda$ does not have simple poles results in the system of algebraic equations
\begin{equation}
\Phi'(z_i)+\sum_{i=1, i\not=j}^{N}\frac{{\cal Q}_j}{z_i-z_j}=0, \quad i=1\dots N.
\label{external}
\end{equation}
Equations (\ref{external}) are the equilibrium conditions for charges in an external harmonic electric field\footnote{Such conditions were first presented by Stieltjes \cite{St,Szego} for roots of orthogonal polynomials.}, or equilibrium of vortices in an irrotational background flow, see Section \ref{vortex}. The function $\Phi(z)$ is a potential of the external field/background flow. Provided (\ref{external}) holds
$$
u(z)=\sum_i\frac{{\cal Q}_i({\cal Q}_i-1)}{(z-z_i)^2}+\Phi''(z)+\Phi'(z)^2+2\sum_i {\cal Q}_i\frac{\Phi'(z)-\Phi'(z_i)}{z-z_i}-\lambda.
$$
If there exists another quasi-rational eigenfunction of $H$ which is of the form similar to (\ref{psi2})
$$
\tilde\psi(z)=\prod_{i=1}^{\tilde N}(z-\tilde z_i)^{{\cal \tilde Q}_i} e^{\Phi(z)}, \quad H\tilde\psi=\eta\tilde\psi,
$$
it can be taken as a ``seed" for the Darboux transformation. Then the transform $\hat\psi$ of $\psi$
\begin{equation}
\hat\psi=\psi'-\frac{\tilde\psi'}{\tilde\psi}\psi, \quad \hat H\hat\psi=(-\partial_z^2+\hat u)\hat\psi=\lambda\hat\psi
\label{tildehatpsi}
\end{equation}
will be also quasi-rational of the form similar to (\ref{psi2}), i.e.,
$$
\hat\psi(z)=C\prod_{i=1}^{\hat N}(z-\hat z_i)^{{\cal \hat Q}_i} e^{\Phi(z)}.
$$
Since the transform of potential
$$
\hat u=u-2(\log\tilde\psi)''
$$
does not have first-order poles, $\hat\psi$ corresponds to a new equilibrium configuration in the same external field.

Thus, we can produce various equilibrium configurations by applying sequence of Darboux transformations to a set of eigenfunctions of $H$ of the form
$$
\varkappa_i=R_i(z)e^{\Phi(z)}, \quad H\varkappa_i=\lambda_i\varkappa_i,
$$
where $R_i(z)$ are rational in $z$. According to the Darboux-Crum theorem (\ref{dtNpsi},\ref{dtNu}), the results of compositions of $n$ transformations can be written in the Wronskian form
\begin{equation}
\psi^{(n)}(z)=\frac{\Wr [\varkappa_{k_1}, \varkappa_{k_2}, \dots, \varkappa_{k_n}]}{\Wr [\varkappa_{k_1}, \varkappa_{k_2}, \dots, \varkappa_{k_{n-1}}]}.
\label{psimany}
\end{equation}
For example, for a harmonic oscillator $u=z^2/2$, $\Phi(z)=-z^2/2$, and $R_n(z)$ is the $(n-1)$th Hermite polynomial. In this case, transforms (\ref{psimany}) correspond to various equilibria of vortices/charges in the quadrupole background flow/external field (for details and  other examples, see e.g. \cite{Cl,Lbhe}). Note that, in contrast to the systems considered in the previous sections, configurations corresponding to Hermite polynomials do not depend non-trivially on continuous parameters and are rather parametrised by sequences of distinct integers $0<k_1<k_2<\dots<k_{n-1}$. These correspond to $\eta_0=\lambda_{k_1},\eta_1=\lambda_{k_2}, \dots, \eta_{n-1}=\lambda_{k_n}$ in the chain (\ref{dtchain}).

\section{Equilibria of Vortex Streets - Odd Family}
\label{streets}

A periodic vortex street is a vortex pattern consisting of an infinite number of vortices situated periodically along some direction. A vortex street can be equivalently considered as a configuration of $N$ vortices (or charges) on a (flat) cylinder. In what follows, the terms ``vortex street", ``periodic configuration" or ``periodic pattern" refer to a configuration of a finite number of charges on cylinder. Without loss of generality, we can set the translation period to $\pi$. Since $ \sum_{n=-\infty}^\infty\frac{1}{z+\pi n}=\cot z $, the stationarity condition (\ref{stationary}) for the translating (with a speed ${\rm i}\bar k/(2\pi)$) vortex street becomes
\begin{equation}
k+\sum_{j=1, j\not=i}^N {\cal Q}_j\cot(z_i-z_j)=0, \quad i=1\dots N ,
\label{stationarypi}
\end{equation}
and for static periodic patterns (see (\ref{static})), we have
\begin{equation}
\sum_{j=1, j\not=i}^N {\cal Q}_j\cot(z_i-z_j)=0, \quad i=1\dots N .
\label{staticpi}
\end{equation}
Similarly to vortex patterns on the plane, which are related to the rational solutions of the KdV hierarchy or the odd bispectral family, the trigonometric soliton solutions of the hierarchy are related to the patterns on a cylinder \cite{HV,KC,Lbhe}. This can be demonstrated by methods similar to those employed in Sections \ref{dtequilibria}, \ref{translating}.

In more detail, we introduce the trigonometric-rational analog of (\ref{psiQ}) of the following form
\begin{equation}
\psi(z)=C\prod_{i=1}^N\sin^{{\cal Q}_i}(z-z_i),
\label{psiQpi}
\end{equation}
where ${\cal Q}_i$, $z_i$ correspond to the periodic static configuration (\ref{staticpi}). The ``trigonometric-rational" here means that all ${\cal Q}_i$ are integers. Trigonometric-rational functions are special cases of the trigonometric-factorizable functions, where ${\cal Q}_i$ are real. Function (\ref{psiQpi}) is an eigenfunction of the Schrodinger operator
$$
H\psi=\lambda\psi, \quad H=-\partial_z^2+u(z)
$$
with the potential
$$
u(z)=\frac{\psi''(z)}{\psi(z)}+\lambda .
$$
Taking into account the identity
$$
\cot(z-z_i)\cot(z-z_j)=-1+\left(\cot(z-z_i)-\cot(z-z_j)\right)\cot(z_i-z_j) ,
$$
similarly to the rational case, we come to the conclusion that the static equilibrium condition is equivalent to the absence of first-order poles in the potential and
\begin{equation}
u=\sum_{i=1}^N \frac{{\cal Q}_i({\cal Q}_i-1)}{\sin^2(z-z_i)}+{\rm const}.
\label{upi}
\end{equation}
A family of vortex street configurations can be generated by the chain of Darboux transformations (\ref{dtchain}) with $\eta_{i-1}=k_i^2$, where $0<k_1<k_2<\dots<k_i<\dots$ is a sequence of positive integers. The chain begins with the free Schrodinger operator $H_0=-\partial_z^2$, and is constructed using ``seed functions" that are its periodic eigenfunctions:
\begin{equation}
\varkappa_{k_i}(z)=\sin (k_i z+\zeta_i).
\label{varkappai}
\end{equation}
Then, according to the Crum theorem (\ref{dtNpsi},\ref{dtNu}), at $n$th step of the chain we have
\begin{equation}
\psi^{(n)}(z)=\frac{\tau_{n+1}(z)}{\tau_n(z)}, \quad \tau_n={\cal W}\left[\varkappa_{k_1}, \varkappa_{k_2}, \dots, \varkappa_{k_n}\right], \quad \tau_0=1 .
\label{psinpi}
\end{equation}
Clearly, $\tau_n(z)$ are trigonometric polynomials, $\psi^{(n)}(z)$ are trigonometric-rational functions and corresponding potentials
\begin{equation}
u_n(z)=-2(\log\tau_n(z))''
\label{un}
\end{equation}
are of the form (\ref{upi}) and do not have simple poles. Thus, the eigenfunctions (\ref{psinpi}) correspond to static vortex street configurations. Note that the potentials (\ref{un}) are periodic $n$-soliton solutions of the KdV hierarchy, and $\tau_n$ are its $n$-soliton $\tau$-functions (see, e.g., \cite{MS}).  We will demonstrate in Section \ref{sources} that this set of configurations actually includes all static vortex street patterns with ${\cal Q}_i=\pm 1$.

For translating configurations (see (\ref{stationarypi})) related to the soliton solutions, we can apply method similar to that used in section \ref{translating}. By analogy with quasi-rational (\ref{psik}), we now consider the trigonometric quasi-rational eigenfunctions:
\begin{equation}
\psi(k,z)=C\prod_{i=1}^N\sin(z-z_i(k))^{{\cal Q}_i}e^{kz}, \quad {\cal Q}_i \in \mathbb{Z} .
\label{psikpi}
\end{equation}
According to the Darboux-Crum theorem (\ref{dtNpsi}), the eigenfunctions obtained through a sequence of the Darboux transformation from the eigenfunction $e^{kz}$ of the free Schrodinger operator are given by:
\begin{equation}
\psi^{(n)}(k,z)=\frac{{\cal W}\left[\varkappa_{k_1}, \varkappa_{k_2}, \dots, \varkappa_{k_n},e^{kz}\right]}{{\cal W}\left[\varkappa_{k_1}, \varkappa_{k_2}, \dots, \varkappa_{k_n}\right]},
\label{BAs}
\end{equation}
where $\varkappa_{k_i}$ are given by (\ref{varkappai}). Obviously, $\psi^{(n)}(k,z)$ are of the form (\ref{psikpi}) and, therefore, correspond to the translating vortex street configurations. Since each $\varkappa_{k_i}$ (\ref{varkappai}) depends on a free parameter, both the translating and static configurations related to soliton solutions depend on $n$ free parameters as well as on the sequence of integers $0<k_1<k_2\dots<k_n$. Up to a $z$-independent common factor $C(k)$, (\ref{BAs}) is the trigonometric Baker-Akhieser function of the KdV hierarchy.

The class of configurations determined by (\ref{varkappai},\ref{BAs}) includes all possible translating vortex street patterns with ${\cal Q}_i=\pm 1$, as will be shown in Section \ref{sources}.

\section{Equilibria of Vortex Streets - Even Family}
\label{streetseven}

In the previous section, we discussed vortex street configurations that are periodic generalizations of those related to the odd family of bi-spectral Schrödinger operators. Below, we introduce configurations that are periodic generalizations of those related to the even family (see section \ref{Bispectral}).

We first consider a chain of Darboux transformations that begins with $H_0=-\partial_z^2+u_0$, where $u_0$ is the trigonometric Poschl-Teller potential
\begin{equation}
u_0=\frac{a(a-1)}{\sin^2 z}+\frac{b(b-1)}{\cos^2 z},
\label{PT}
\end{equation}
and $a$ and $b$ are arbitrary real numbers. The simplest configuration corresponds to the eigenfunction $\sin^a z\cos^b z$ and consists of two static vortices of with strengths ${\cal Q}_1=a$, ${\cal Q}_2=b$ situated at diametrically opposite points of the $\pi$-periodic cylinder, i.e., at $z=z_1=0$ and $z=z_2=\pi/2$.

When $a$ and $b$ are integers, the potential (\ref{PT}) is a soliton solution of the KdV hierarchy at specific ``times" (the corresponding parameters $k_i$ and $\zeta_i$ in (\ref{psinpi}, \ref{varkappai}) can be found, e.g., in refs \cite{BL} or \cite{GM})\footnote{In general, a sequence of $n$ Darboux transformations from the chain (\ref{dtchain}), where $A_i=\partial_z-(a+i)\cot z+(b+i)\tan z$, $i=0,1,\dots$ , results in a shift of parameters $a\to a+n$, $b\to b+n$ in the potential (\ref{PT}).}. In this case, $u_0$ and sequences of its Darboux transforms belong to the odd family considered in the previous section.

Due to invariance of the potential (\ref{PT}) under the involutions $a\to 1-a$, $b\to 1-b$, for generic $a$ and $b$ (i.e., $a\pm b\not\in\mathbb{Z}$), the Schrodinger operator $H_0=-\partial_z^2+u_0$ has four distinct types of factorizable eigenfunctions $\varkappa_i(z)$, $H_0\varkappa_i=\lambda_i\varkappa_i$ of the form :
\begin{equation}
\begin{array}{ll}
{}_2F_1(-i,i+a+b,1/2+a;\sin^2 z)\sin^a z\cos^b z, & \lambda_i=(2i+a+b)^2 ,\\
{}_2F_1(-i,i+1-a+b,3/2-a\;\sin^2 z)\sin^{1-a} z\cos^b z, & \lambda_i=(2i+1-a+b)^2 , \\
{}_2F_1(-i,i+a+1-b,1/2+a;\sin^2 z)\sin^a z\cos^{1-b} z, & \lambda_i=(2i+1+a-b)^2 , \\
{}_2F_1(-i,i+2-a-b,3/2-a;\sin^2 z)\sin^{1-a} z\cos^{1-b} z, & \lambda_i=(2i+2-a-b)^2,
\end{array}
\label{varkappaab}
\end{equation}
where $i=0,1,2, \dots $ and ${}_2F_1$ stands for the Gauss hypergeometric function. The first parameter of the ${}_2F_1$  in (\ref{varkappaab}) is a non-positive integer $-i$, and the hypergeometric factors in (\ref{varkappaab}) are polynomials in $\sin^2 z$. These factorizable eigenfunctions are trigonometric-rational functions, up to a factor of $\sin^{\pm a}z\cos^{\pm b}z$. For generic $a$ and $b$, all eigenvalues $\lambda_i$ are distinct, and all eigenfunctions in (\ref{varkappaab}) are linearly independent. 

From (\ref{varkappaab}), using the Wronskian formula (\ref{psinpi}), we can construct eigenfunctions of Darboux transforms $H_n=-\partial_z^2+u_0-2(\log\tau_n)''$ of $H_0$. These eigenfunctions are factorizable, and, therefore, correspond to static vortex street configurations. Note that, for generic $a$ and $b$, only one of the two linearly independent solutions of the Schrodinger equation $H_0\varkappa_i=\lambda_i\varkappa_i$ is factorizable. Therefore, in contrast to the soliton-related configurations (\ref{varkappai},\ref{psinpi}), the configurations generated from the seed functions (\ref{varkappaab}) do not depend on non-trivial free parameters. Similar to the systems considered in Section \ref{nonuniform}, configurations with generic $a$ and $b$ are parametrized only by a sequence of distinct integers $k_1<k_2<\dots<k_n$. These correspond to $\eta_0,\eta_1,\dots,\eta_{n-1}$ in the chain (\ref{dtchain}), where each $\eta_i$ belongs to the set of eigenvalues given in (\ref{varkappaab}) . Here, generic configurations consist of four species of vortices: two species with ${\cal  Q}_i=\pm 1$ at $z=z_i\not\in\{0,\pi/2\}$, a vortex of the third species at $z=0$, and a vortex of the fourth species at $z=\pi/2$.

An even family of periodic solutions emerges when $a$ and $b$ become half-integers, such that
$$
a=l+1/2, \quad b=m+1/2, \quad l,m \in \mathbb{N} .
$$
In this case, eigenvalues $\lambda_i$ in (\ref{varkappaab}) overlap, and for some of them ``degeneration" occurs\footnote{Degeneration also occurs for integer $a$ and $b$. There, number of degenerate factorizable eigenfunctions, as we saw in the previous section, is infinite.}. Now, there are two types (degenerate and non-degenerate) of factorizable eigenfunctions $\varkappa_i(z)$:
\begin{equation}
\begin{array}{lll}
P^{(-l,-m)}_i(\cos 2z, \zeta_i)\sin^{1/2-l} z\cos^{1/2-m} z, & \lambda_i=(2i+1-l-m)^2 , & \frac{l+m}{2} \le i < l+m  \\
P^{(l,m)}_i(\cos 2z)\sin^{l+1/2} z\cos^{m+1/2} z, & \lambda_i=(2i+1+l+m)^2 , & i\ge 0
\end{array} .
\label{varkappahalf}
\end{equation}
These are both trigonometric-rational up to the common factor $(\sin z\cos z)^{1/2}$. In the first type, $P^{(-l,-m)}_i(x,\zeta)$ is a polynomial in $x$ that also depends on a free parameter $\zeta$. Such polynomials are called para-Jacobi Polynomials \cite{BGQ,CY}. Solutions of the second type involve ordinary Jacobi polynomials $P^{(l,m)}_i$ and are non-degenerate.

Thus, the even family of the static vortex street configurations parametrized by sequences of continuous free parameters as well as sequences of integers can be constructed with the help of (\ref{psinpi}) and (\ref{varkappahalf}) for half-integer $a$, $b$.

When $l\not=m$ (i.e., $a\not=b$), the generic configurations consist of four species of vortices: two species with ${\cal  Q}_i=\pm 1$ at $z=z_i\not\in\{0,\pi/2\}$, a vortex of the third species at $z=0$, a vortex of the fourth species at $z=\pi/2$.

When $l=m$ (i.e., $a=b$), the vortices at $z=0$ and $z=\pi/2$ have the same strengths, and the generic $l=m$ configurations consist of three species of vortices.

The simplest non-trivial example of a member of the even family corresponds to $l=m=1$. Here,  $a=b=3/2$, and $P^{(-1,-1)}_1(\cos 2z, \zeta_1)=\cos 2z+1+\zeta_1$. From (\ref{varkappahalf}), we obtain
$$
\psi_1=\varkappa_1=\frac{\cos 2z+1+\zeta_1}{(\sin z\cos z)^{1/2}}.
$$
Note that by the scaling $z\to \epsilon z$, $\psi_1\to -\psi_1/(2\epsilon^{3/2})$, and by the change of parameter $\zeta_1=-2(1+\epsilon^2 s_1)$, in the $\epsilon\to 0$ limit, we obtain
$$
\psi_1\to\lim_{\epsilon\to 0}\frac{-1}{2\epsilon^{3/2}}\sqrt{2}\frac{\cos (2\epsilon z)+1-2(1+\epsilon^2 s_1)}{\sqrt{\sin(2 \epsilon z)}}=\frac{z^2+s_1^2}{z^{1/2}} .
$$
This is $\psi_1$ (see (\ref{varkappa01})) from the even bi-spectral family discussed in Section \ref{Bispectral}.

In general, the terminating sequences described in Section \ref{Bispectral} are the rational (times $z^{1/2}$) limits of the trigonometric sequences of the case $a=b=m+1/2$
\begin{equation}
\psi_m={\cal W}[\varkappa_m], \quad \psi_{m+1}=\frac{{\cal W}[\varkappa_m,\varkappa_{m+1}]}{{\cal W}[\varkappa_m]}, \dots, \psi_{2m-1}=\frac{{\cal W}[\varkappa_m,\varkappa_{m+1}, \dots, \varkappa_{2m-2}, \varkappa_{2m-1}]}{{\cal W}[\varkappa_m,\varkappa_{m+1}, \dots, \varkappa_{2m-2}]} , 
\label{evenpi}
\end{equation}
where
\begin{equation}
\varkappa_{m+i}=(\sin z\cos z)^{1/2-m}P^{(-m,-m)}_{m+i}(\cos 2z, \zeta_{m+i}), \quad i=0, 1, \dots, m-1 .
\label{varkappaeven}
\end{equation}
Here, polynomials $P^{(-m,-m)}_n$ are the special $l=m$ case of the para-Jacobi polynomials (``para-Gegenbauer" polynomials)\footnote{They satisfy equation $\left((1-x^2)\partial_x^2+2(m-1)x\partial_x+n(n-2m+1)\right)P^{(-m,-m)}_n(x,\zeta)=0$, which reduces to the standard hypergeometric form under the change of variable $w=(1-x)/2$.}
$$
\begin{array}{l}
P^{(-m,-m)}_n(x,\zeta)=\frac{(-2)^n(n-m)!n!}{(2n-2m)!}
\sum_{k=0}^{n-m}\frac{(-1)^{n-k}(2n-2m-k)!}{k!(n-m-k)!(n-k)!}\left(\frac{1+x}{2}\right)^{n-k} + \\
\qquad\qquad\qquad\qquad
\zeta\frac{(-2)^n(2n-2m+1)!(2m-n-1)!}{(n-m)!}
\sum_{2(n-m)+1}^n \frac {(-1)^{n-k}(k-n+m-1)!}{k!( k+2m-2n-1)!(n-k)!}\left(\frac{1+x}{2}\right)^{n-k} .
\end{array}
$$
When we re-scale $z\to\epsilon z$, $H \to \epsilon^2 H$, the eigenvalues scale as $\lambda_i\to\epsilon^2\lambda_i$, and in the rational limit, we have confluent eigenvalues, all tending to zero: $\lambda_i\to\eta=0$. According to Section \ref{DTFC} (see eqs. (\ref{WXN}), (\ref{H0Xn})), in this limit, the Wronskian representation of the Darboux transforms becomes
\begin{equation}
\psi_{m+i}=\frac{\tau_{i+1}}{\tau_i}, \quad \tau_i=\Wr[{\cal X}_1, {\cal X}_2, \dots, {\cal X}_i] , \quad \tau_0=1 ,
\label{tauiX}
\end{equation}
where now
$$
H_0{\cal X}_1(z)=0, \quad H_0{\cal X}_{i+1}(z)=c_i{\cal X}_i(z), \quad i=1 \dots m-1.
$$
Here, $H_0=-\partial_z^2+\frac{m^2-1/4}{z^2}$ is the rational limit of the trigonometric Schrodinger operator (\ref{PT}), with $a=b=m+1/2$, obtained by the scaling $z\to\epsilon z$, $H_0\to\epsilon^2H_0$, i.e. $H_0\to\lim_{\epsilon=0}\epsilon^2\left(-\partial^2/\partial(\epsilon z)^2+u_0(\epsilon z)\right)$. In other words, ${\cal X}_i(z)$ are Laurent polynomials in $z^{1/2}$ that solve the following chain of equations:
\begin{equation}
{\cal X}_1=z^{1/2-m}(z^{2m}+s_m) , \quad -{\cal X}_{i+1}''+\frac{m^2-1/4}{z^2}{\cal X}_{i+1}=c_i{\cal X}_i, \quad i=1,2, \dots, m-1 ,
\label{XNeven}
\end{equation}
where $c_i$ are arbitrary constants. Although two constants of integration appear at each step of the chain, only one of them is essential. The chain terminates at the $(m-1)$th step, since the logarithmic term appears at the $m$th step. The function ${\cal X}_i$ is the $\epsilon\to 0$ limit of a linear combinations of $i$ first eigenfunctions (\ref{varkappaeven}):
$$
\epsilon^{3/2-2i-m}\varkappa_m(\epsilon z,\zeta_m),\quad \epsilon^{3/2-2i-m}\varkappa_{m+1}(\epsilon z,\zeta_{m+1}), \dots , \epsilon^{3/2-2i-m}\varkappa_{m+i-1}(\epsilon z,\zeta_{m+i}) ,
$$
where the $(j-1)$th free parameter $\zeta_{m+j}$ is a linear combination of
$$
1, \quad \epsilon^{2m}s_m, \quad \epsilon^{2m+2}s_{m+1},  \quad \epsilon^{2m+4}s_{m+1}, \dots , \epsilon^{2(m+j-1)}s_{m+j} .
$$
The coefficients of the linear combinations are constants determined by the condition of the existence of the limit.

For example, in the case $m=2$, corresponding to the terminating sequence (\ref{P3P4}), we have
$$
\begin{array}{l}
{\cal X}_1=\lim_{\epsilon\to 0} \epsilon^{-5/2}\varkappa_2(\epsilon z, 2-2\epsilon^4 s_2), \\
{\cal X}_2=\lim_{\epsilon\to 0} \epsilon^{-9/2}\left(\varkappa_3\left(\epsilon z, -\frac{1}{2}-\frac{3}{2}\epsilon^4s_2-\frac{1}{2}\epsilon^6 s_3\right)-3\varkappa_2(\epsilon z, 2-2\epsilon^4 s_2)
\right) .
\end{array}
$$
Concluding this section we note that, in the trigonometric case, sequences of the static periodic configurations (\ref{evenpi}) can be continued with the help of solutions involving the ordinary Jacobi polynomials (see (\ref{varkappahalf})). In other words, unlike the rational limit, trigonometric sequences are not terminating. However, similar to the rational case, these sequences depend only on a finite number of free continuous parameters.

\section{Time-Dependent Darboux Transformations, Calogero-Moser Systems and Locus Configurations}.
\label{sources}

As we saw in Section \ref{L12}, all possible static configurations of a finite number of vortices of strengths ${\cal Q}_i=\pm 1$ are given by roots of the Adler-Moser polynomials. In this section we demonstrate that sequences of soliton $\tau$-functions (\ref{varkappai},\ref{psinpi}) define all possible vortex street patterns with  ${\cal Q}_i=\pm 1$. To this end, we introduce dynamical systems that are completely integrable in the $\Lambda=1$ case, and whose fixed points coincide with equilibrium configurations of the point vortices. These systems are related to Darboux transforms of the time-dependent free Schrodinger (or heat) equation.

In more detail, we consider the imaginary-time Schrodinger equation with time-dependent potential
\begin{equation}
\frac{\partial\psi(z,t)}{\partial t}=H\psi(z,t), \quad H=-\partial_z^2+u(z,t).
\label{tSE}
\end{equation}
Here, $H$ can be expressed in the form:
\begin{equation}
H=A^*A+(\log\varkappa)_t \,,
\label{AplusAt}
\end{equation}
where subscript $t$ denotes the time derivative. In equation (\ref{AplusAt}), $\varkappa=\varkappa(z,t)$ is one of solutions of (\ref{tSE}). The first-order differential operators $A$ and $A^*$ are of the same form as in equation (\ref{AstarA}). Permuting factors in (\ref{AplusAt}), we obtain the new Schrodinger operator
$$
\hat H=AA^*+(\log\varkappa)_t .
$$
Multiplying (\ref{AplusAt}) by $A$ from the left, we obtain $AH=\hat H A-A_t$, and therefore $\hat\psi=A\psi$ solves the new equation
$$
\frac{\partial\hat\psi}{\partial t}=\hat H\hat\psi, \quad \hat H=-\partial_z^2+\hat u(z,t).
$$
The new potential $\hat u(x,t)$ and solution $\hat\psi$ are given by the same formulae as in the case of the time-independent Darboux transformations, i.e.,
\begin{equation}
\hat\psi=A\psi=\psi'-(\log\varkappa)'\psi, \quad \hat u = u -2(\log \varkappa)'' .
\label{dt}
\end{equation}
Now, we take the free Schrodinger operator $H_0=-\partial_z^2$. The corresponding imaginary-time Schrodinger (or inverse heat) equation admits polynomial solutions $q(z,t)$:
$$
q_t=-q'', \quad q=\prod_{i=1}^l(z-z_i(t)) ,
$$
where $z_i(t)$ obey the dynamical equations
$$
\frac{d z_i}{dt}=\sum_{j=1, j\not=i}^l\frac{2}{z_i-z_j}, \quad i=1\dots l .
$$
Using $q$ as a seed function ($\varkappa=q$ in Eq. (\ref{dt})) we can apply the Darboux transformation to another polynomial solution $\psi$, $\psi_t=-\psi''$. The transform $\hat\psi$ of $\psi$ is a rational function:
\begin{equation}
\hat\psi=\frac{p}{q}, \quad p=\psi'q-q'\psi.
\label{psiqp}
\end{equation}
The Schrodinger operator acquires the potential $\hat u=-2(\log p)''$, and the new equation $\hat H\hat\psi=\hat\psi_t$ writes, in terms of $p$ and $q$, as
$$
\left(-\partial_z^2-2(\log q)''\right)\left[p/q\right]=\left[p/q\right]_t ,
$$
or equivalently
$$
pq_t-qp_t=p''q-2p'q'+q''p.
$$
Thus, we have incorporated dynamics into the Tkachenko equation (\ref{Tkachenko}). It is easy to see that further iterations of Darboux transformations will lead to equations of the same form for consecutive transforms of polynomials.

The time-dependent Tkachenko equation results in the following dynamical system for the roots of $p$ and $q$, denoted as $z_1, \dots, z_l$ and $z_{l+1}, \dots, z_{l+m}$ respectively:
\begin{equation}
\frac{dz_i}{dt}=2\sum_{j=1, j\not=i}^N\frac{{\cal Q}_j}{z_i-z_j}, \quad i=1\dots N , \quad N=l+m,
\label{chudnovskij}
\end{equation}
where negative and positive charges of equal magnitude ${\cal Q}_i=\pm 1$ are attached to roots of $p$ and $q$, respectively.

System (\ref{chudnovskij}), with arbitrary\footnote{This is a system for the roots of polynomials ${\cal P}_i(z,t)$, which satisfy the time-dependent generalization of the poly-linear equation (\ref{multilinear}): $-\sum_i\Lambda_i\frac{\partial {\cal P}_i/\partial t}{{\cal P}_i}=\sum_i\Lambda_i^2\frac{{\cal P}_i''}{{\cal P}_i}+2\sum_{i<j}\Lambda_i\Lambda_j\frac{{\cal P}_i'}{{\cal P}_i}\frac{{\cal P}_j'}{{\cal P}_j}$, see \cite{Lbhe}.} ${\cal Q}_i$, can be embedded into a hamiltonian system of newtonian particles interacting pairwise through inverse square potentials: By taking the time derivative of eq. (\ref{chudnovskij}), and then eliminating the first order time derivatives by substituting them from (\ref{chudnovskij}), we obtain:
\begin{equation}
\frac{d^2z_i}{dt^2}=-4\sum_{j=1,j\not=i}^N\frac{{\cal Q}_j({\cal Q}_i+{\cal Q}_j)}{(z_i-z_j)^3}
\label{LCM}
\end{equation}
(For details, see e.g. \cite{Lbhe} or \cite{S}). For $\Lambda=1$, the sum of two charges vanishes, ${\cal Q}_i+{\cal Q}_j=0$, when $i$ and $j$ belong to different species, and (\ref{LCM}) decouples into two non-interacting sub-systems
$$
\frac{d^2z_i}{dt^2}=-\sum_{j:j\not=i}\frac{8}{(z_i-z_j)^3},
$$
one consisting of charges ${\cal Q}_i=-1$ at $z_1, \dots, z_l$ and another with ${\cal Q}_i=1$ at $z_{l+1}, \dots , z_{l+m}$. These are completely integrable Calogero-Moser systems \cite{Moser}.

There is another case of decoupling, this time for three species of symmetrically situated particles: One particle of strength $a$ at the origin $z=0$, $2l$ particles of strength $-1$ at $\pm z_1, \dots, \pm z_l$ and $2m$ particles of strength $1$ at $\pm z_{l+1}, \dots , \pm z_{l+m}$. Here, due to the symmetry, the system reduces from $2(l+m)+1$ to $l+m$ degrees of freedom and decoupling into the two BC-type Calogero-Moser subsystems
$$
\frac{d^2z_i}{dt^2}=\frac{1-(2a\pm 1)^2}{z_i^3}-\sum_{j:j\not=i}\left(\frac{8}{(z_i-z_j)^3}+\frac{8}{(z_i+z_j)^3}\right)
$$
of $l$ and $m$ particles respectively\footnote{For a review of Calogero-Moser systems related to different root systems, see e.g. \cite{OP}.}. This case corresponds to the Darboux-transformed imaginary-time Schrodinger equation with
$$
H_0=-\partial_z^2+\frac{a(a-1)}{z^2}.
$$
The initial equation $H_0\psi=\psi_t$ has quasi-polynomial solutions of the form $\psi=z^a\prod_{i=1}^l(z^2-z_i(t)^2)$, and the Darboux transforms constructed from these solutions are quasi-rational. When $a$ is a half-integer, the fixed points of system (\ref{chudnovskij}) correspond to configurations of the even bi-spectral family from Section \ref{Bispectral}.

Let us now return to the main topic of this section and consider equilibrium configurations of vortices. For these configurations $d^2z_i/dt^2=0$. Then form (\ref{LCM}) it follows that the equilibrium conditions (\ref{static}) imply the following $N$ locus conditions\footnote{Note that additional locus conditions for the case ${\cal Q}_i\in\{-1,2\}$ were found by O'Neil \cite{ON3}.}
$$
\sum_{j=1,j\not=i}^N\frac{{\cal Q}_j({\cal Q}_i+{\cal Q}_j)}{(z_i-z_j)^3}=0, \quad i=1\dots N,
$$
i.e., the locus conditions are necessary conditions for equilibrium. In the case of two species with ${\cal Q}_i=\pm 1$, locus conditions decouple, resulting in a separate set of locus conditions\footnote{It is worth mentioning that the theory of algebraically integrable systems also considers loci with multiple roots, where more general locus conditions are imposed \cite{Chalykh,CFV2}.} for each species:
\begin{equation}
\sum_{j:j\not=i}\frac{1}{(z_i-z_j)^3}=0  ,
\label{locus}
\end{equation}
with $l$ conditions for the first species and $m$ conditions for the second species respectively.

It turned out \cite{AirMcKeMos}\footnote{Conditions (\ref{locus}) first appeared in this famous paper by Airault, McKean
and Moser, who introduced the term ``locus".} that set of solutions to (\ref{locus}) is not empty only if the number of particles the locus is formed of is a triangular number $n(n+1)/2$. Subsequently, it was shown \cite{AM} that for this number, roots of the Adler-Moser polynomials $P_n$ constitute complete solution of (\ref{locus}). As we saw in Section \ref{L12}, in systems of vortices in static equilibrium,  $l=n(n-1)/2$ and $m=n(n+1)/2$ particles form two loci that are roots of two consecutive Adler-Moser polynomials $P_{n-1}(z,s_1, \dots, s_{n-2})$ and $P_n(z,s_1, \dots, s_{n-1})$.

Conditions (\ref{locus}) must also hold for each species of translating configurations (\ref{stationary}) with ${\cal Q}_i=\pm 1$. Recall that, due to neutrality condition (\ref{neutrality}), for translating configurations $\deg p=\deg q$, i.e., $l=m$. This means that when $k\not=0$, both $p$ and $q$ are Adler-Moser polynomials of the same degree, but with different sets of parameters. As a result, according to Section \ref{translating}, all possible translating configurations consisting of a finite number of vortices with ${\cal Q}_i=\pm 1$ are given by (\ref{AMtranslating}). Specifically, $q=P_n(z,s_1,\dots,s_{n-1})$, $p=P_n(z+\tilde s_0, \tilde s_1, \dots, \tilde s_{n-1})$, where $\tilde s_0, \dots, \tilde s_{n-1}$ are functions of $s_1,\dots,s_{n-1}$ and $k$.

In the trigonometric (periodic) case, for each of two species of opposite charges the locus conditions are writen as
$$
\sum_{j:j\not=i}\frac{\cos(z_i-z_j)}{\sin^3(z_i-z_j)}=0 .
$$
It is known (see e.g. \cite{Chalykh,CFV2}) that all trigonometric locus configurations are given by the roots of soliton $\tau$-functions of the KdV hierarchy. Therefore, any pair $p$ and $q$ that does not share common roots and satisfies the periodic Tkachenko equation
\begin{equation}
p''q-2p'q'+pq''+\left(l-m\right)^2pq=0, \quad p=\prod_{i=1}^l\sin(z-z_i), \quad q=\prod_{i=1}^m\sin (z-z_{l+i}) 
\label{Tkachenkopi}
\end{equation}
consists of soliton $\tau$-functions. On the other hand, such $\tau$-functions determine potentials in the factorization chain (\ref{dtchain}), which begins with the free Schrodinger operator. In this chain of transformations
$$
H_n\psi^{(n)}=\eta_n\psi^{(n)}, \quad H_n=-\partial_z^2-2(\log \tau_n)'', \quad \psi^{(n)}=\frac{\tau_{n+1}}{\tau_n}, \quad \tau_0=1,
$$
i.e.,
\begin{equation}
\left(-\partial_z^2-2(\log \tau_n)''\right)\left[\frac{\tau_{n+1}}{\tau_n}\right]=\eta_n \frac{\tau_{n+1}}{\tau_n} .
\label{Setaupi}
\end{equation}
Since the degree of the leading term of the Laurent expansion of $\tau_{n+1}/\tau_n$ in $\exp({\rm i}z)$ equals $d_{n+1}-d_n$, where $d_n=\deg\tau_n$, we can conclude from (\ref{Setaupi}) that $\eta_n=(d_{n+1}-d_n)^2$. Then, rewriting (\ref{Setaupi}) in the bi-linear form, we obtain the Darboux chain for $\tau_n$:
\begin{equation}
\tau_{n+1}''\tau_n-2\tau_{n+1}'\tau_n'+\tau_{n+1}\tau_n''+(d_{n+1}-d_n)^2\tau_{n+1}\tau_n=0, \quad d_n=\deg \tau_n, \quad \tau_0=1 .
\label{harmonicphi}
\end{equation}
The periodic analog of the Tkachenko equation (\ref{Tkachenkopi}) has the same form as the Darboux chain equation (\ref{harmonicphi}). Taking this into account, along with the fact that $p$ and $q$ are necessarily soliton $\tau$-functions (as follows from locus conditions), we conclude that $p$ and $q$ are soliton $\tau$-functions related by a Darboux transformation. Therefore, all possible static street patterns consisting of vortices of two species of opposite strengths are defined by (\ref{varkappai},\ref{psinpi}).

For translating vortex streets, we use arguments similar to those of the rational case, which leads us to the conclusion that the complete set of translating street patterns with ${\cal Q}_i=\pm 1$ is determined by (\ref{varkappai},\ref{BAs}). It is worth mentioning that, according to the soliton theory, the following relationship between $p$ and $q$ holds
$$
q=\tau_n(z, \sigma_1, \dots, \sigma_n), \quad p=\tau_n(z, \zeta_1, \dots, \zeta_n) , \quad (k+{\rm i} k_j)e^{2{\rm i}\sigma_j}=(k-{\rm i} k_j)e^{2{\rm i}\zeta_j} ,
$$
where $\tau_n$ is given by  (\ref{varkappai},\ref{psinpi}). Without going into much detail, we note that this relationship can be derived using the Sato formula for the Baker-Akhieser function of the KP hierarchy, as well as the correspondence between Hirota and Wronskian representations of the $n$-soliton $\tau$-functions (for an introduction to the soliton theory, see e.g. \cite{Hirota,Z}).

\section{Conclusions and Open Problems}
\label{harmonic}

In this paper, we reviewed developments in applications of the factorization method to the theory of point vortex patterns that have occurred over the past several decades, and presented new results.  These developments are linked to the theory of hierarchies of integrable PDEs and the bi-spectral problem.

Connection with the KdV hierarchy has been known for a long time. The Lax operator of this hierarchy is a Schrodinger operator, and related non-terminating configurations (i.e., $\Lambda=1$ non terminating sequences), both static and translating, on the plane and on the cylinder, can be generated through the factorization of rational or trigonometric Lax-Schrodinger operators. Multi-parametric terminating static configurations, both on the plane and on the cylinder, can also be generated through the factorization of the Schrodinger operator in the $\Lambda=1$ case.

In the $\Lambda=2$ case, connections to the Sawada-Kotera and Kaup-Kupersmidt hierarchies were first found by Demina and Kudryashov \cite{DK1}. The Lax operators of these hierarchies are of third order, so a natural idea was to consider the Darboux transformations for the third-order operators $L=\partial_z^3-u\partial_z$ \cite{LY1}. The development of this idea was reviewed in Sections \ref{3rd_order} and \ref{3rd_order_equilibrium} (also see Appendix \ref{AA}). The non-terminating and terminating sequences of static configurations on the plane can be generated by such transformations in the $\Lambda=2$ case.

In contrast to the $\Lambda=1$ case, the translating configurations cannot be constructed either through factorization or the $\tau$-function method when $\Lambda=2$. Indeed, in the $\Lambda=1$ casde, the translating configurations are determined by the Baker-Akhiezer function, which is obtained through the action of the intertwining operator on eigenfunctions of the free Lax-Schrodinger operator. However, the Lax operator is of third order in the $\Lambda=2$ (Sawada-Kotera) case. The corresponding third-order equation for the Baker-Akhiezer function of the hierarchy is (see Appendix \ref{AA} for details):
$$
L\Psi(k,z)=k^3\Psi(k,z), \quad L=\partial_z^3+6(\log q(z))''\partial_z, \quad \Psi(k,z)=\frac{\theta(k,z)}{q(z)}e^{kz} .
$$
It cannot, in general, be reduced to the Schrodinger equation
$$
H\phi(k,z)=-k^2\phi(k,z), \quad H=-\partial_z^2-\Lambda(\Lambda+1)(\log q(k,z))'', \quad \phi(k,z)=\frac{p(k,z)}{q(k,z)^\Lambda}e^{kz} ,
$$
which is equivalent to the bilinear equation for translating configurations. Only when $k=0$ does the eigenvalue problem $L\psi=(\partial_z^2-u)\partial_z\psi=k^3\psi$ reduce to solving the Schrodinger equation $(-\partial_z^2+u)\phi=0$, where $\phi$ is the $k\to 0$ limit of $\psi'$:
$$
\phi(z) = \lim_{k\to 0} \frac{\partial \psi(k,z)}{\partial z}, \quad \psi=C_1(k)\Psi(k,z)+C_2(k)\Psi(e^{2{\rm i}\pi/3}k,z)+C_3(k)\Psi(e^{4{\rm i}\pi/3}k,z),
$$
which corresponds to static configurations (Here, suitably chosen $C_i(k)$ ensure finiteness of the limit). As a consequence, translating configurations cannot be generated through Darboux transformations, and one cannot expect to find non-terminating sequences. Thus, the classification of translating configurations remains an open question for $\Lambda=2$.

A similar problem arises when one tries to classify vortex-street patterns in the $\Lambda=2$ case: 
We recall that non-terminating vortex street configurations in the $\Lambda=1$ case (Section \ref{streets}) are generated through a chain of Darboux transformations at distinct non-zero levels. The corresponding eigenfunctions result from the action of the trigonometric intertwining operators on the trigonometric eigenfunctions $\sin(k_i z+\zeta_i)$, $k_i>0$ of the free Schrodinger operator $H_0=-\partial_z^2$. However, in the $\Lambda=2$ case, we have intertwining between third-order operators, rather than between Schrodinger operators. Moreover, the free Lax operator $L_0=\partial_z^3$ does not have trigonometric eigenfunctions. Therefore, one cannot expect to find non-terminating sequences here.

We searched for sequences of configurations on the cylinder by solving the trigonometric analog of the bilinear equation (\ref{bilinear}):
\begin{equation}
\tau_{n-1}''\tau_n-2\gamma_n \tau_{n-1}'\tau_n'+\gamma_n^2\tau_{n-1}\tau_n''+(d_{n-1}-\gamma_nd_n)^2\tau_{n-1}\tau_n=0 , \quad \gamma_n\gamma_{n+1}=1, \quad d_n=\deg\tau_n .
\label{bilinearpi}
\end{equation}
In the $\Lambda=2$ case, we found only short terminating sequences. For example
$$
\tau_0=\xi-\frac{1}{\xi}, \tau_1=\xi^6-4\xi^4+5\xi^2+\frac{5s_1}{\xi^2}-\frac{4s_1}{\xi^4}+\frac{s_1}{\xi^6},\,
\tau_2=\xi^5-5\xi^3-s_2\xi+\frac{s_2}{\xi}-\frac{5s_1}{\xi^3}+\frac{s_1}{\xi^5},
$$
where $\xi=e^{{\rm i}z}$ and $\gamma_0=2$. The above example can be continued for one more step if we impose condition $s_1=s_2^2/5$. This gives another terminating sequence with $\tau_1=\tau_1(\xi,s_2)$, $\tau_2=\tau_2(\xi,s_2)$ and $\tau_3=\tau_3(\xi,s_2,s_3)$ , where $\deg\tau_3=14$. 

Without going into much detail, we note that, in such sequences, $\tau_1$ is a solution to an ODE that can be reduced to a Gauss hypergeometric form\footnote{Alternatively, $\tau_1$ can be obtained through Darboux transformations of the free Schrodinger operator, so that $\tau_1=C_n\Wr[\sin z,\sin(2z),\sin\left(nz+\zeta\right)]/\sin z$, $C_n=-2{\rm i}\exp(-{\rm i}\zeta)/\left((n-1)(n-2)\right)$, and $\deg\tau_1=n+2$, $s_1=\exp(-2{\rm i}\zeta)$.}. Then we can apply KWCC transformations (\ref{KWCCi}) to $\psi_i=\tau_i^{\gamma_i}/\tau_{i-1}$, $i=1,2$, to continue the sequence. This transformation can be applied at most twice. The terminating configurations obtained in this way can depend on no more than two non-trivial free parameters (they depend on two parameters when $\deg\tau_1=6,8,10,\dots$).

Thus, the problem of complete classification of vortex patterns on the cylinder in the $\Lambda=2$ case remains open.

As discussed in Sections \ref{DTFC}, \ref{streetseven}, in the $\Lambda=1$ case, both main and terminating sequences can be expressed in the form of Wronskians. In the $\Lambda=2$ case, configurations of main sequences have Pfaffian representation (see Appendix \ref{AA}). We note that a determinant representation also exists for the $\Lambda=2$ main sequences, due to connections with the Kaup-Kupershmidt hierarchy (see Section 7 of \cite{LY1} for more details). Accordingly, the natural question arises about Pfaffian/determinant representations for the $\Lambda=2$ terminating sequences. To address this question, one would need to obtain the Phaffian/determinant representation using the factorization method, rather than the $\tau$-function approach.

Finally, classification of all stationary vortex patterns and understanding the role of factorization methods in this classification are ultimate questions to address in this subject.

Concluding this review, we would like to mention a related problem from the theory of algebraically integrable systems, as our interest in vortex patterns stems from that theory: In the case of the odd family of periodic configurations, it is convenient to rewrite the bilinear chain (\ref{harmonicphi}) in terms of homogeneous polynomials in two variables, replacing $\tau_n(z)=\prod_{i=1}^{d_n}\sin(z-z_i)$ with
$$
\tau_n(X,Y)=R^{d_n} \tau_n(z)=\prod_{i=1}^{d_n}(Y\cos z_i-X\sin z_i),
$$
where
$$
X=R\cos z, \quad Y=R\sin z.
$$
In the new variables, the chain (\ref{harmonicphi}) takes the following form:
\begin{equation}
\tau_n\Delta \tau_{n+1}-2(\nabla \tau_{n+1}\cdot\nabla\tau_n)+\tau_{n+1}\Delta \tau_n=0, \quad \tau_0=1, \quad \nabla=(\partial_X,\partial_Y) ,
\label{Harmonic}
\end{equation}
where $\Delta=\nabla\cdot\nabla$ and $\nabla$ denote two-dimensional Laplacian and gradient, respectively. This chain is called harmonic \cite{Bb}, because at the first step $\Delta\tau_1=0$, i.e., $\tau_1$ is a harmonic function, and Schrodinger operators corresponding to $\tau_n$ are natural generalizations of the Laplace operator \cite{L,LY}. The harmonic chain can be generalized to any number of dimensions: Non-terminating solutions of (\ref{Harmonic}) are $\tau$-functions for potentials of algebraically integrable Schrodinger operators (see e.g. \cite{Bb}, \cite{Chalykh}). One can view (\ref{Harmonic}) as a generalization of the Darboux chain (\ref{rchain}), that starts from the free Schrodinger operator\footnote{A chain that starts from $u_0\not=0$ has the form $\tau_n\Delta \tau_{n+1}-2(\nabla \tau_{n+1}\cdot\nabla\tau_n)+\tau_{n+1}\Delta \tau_n-u_0\tau_n\tau_{n+1}=0$, $\tau_0=1$.}, to any number of dimensions. In one dimension, the Adler-Moser polynomials constitute complete set of non-terminating solutions of the chain. In two dimensions, the soliton related solutions exhaust all non-terminating solutions in the class of homogeneous polynomials: The family of two-dimensional solutions consists of an infinite number of branches labelled by $0<k_1<k_2<\dots$, rather than a single sequence, as in one dimension. Complete classification of non-terminating solutions in all dimensions is a hard open problem (see e.g. \cite{Bb,Chalykh})\footnote{It is conjectured that families listed in \cite{Chalykh} constitute complete solution of the problem.}.

\begin{appendices}

\section{Main Sequences and Polynomial $\tau$-functions: Wronskian and Pfaffian Representations. }
\label{AA}

The zero-level eigenfunction $\psi$ of a rational Lax operator $L$ of an integrable hierarchy has the form $\psi=\theta/\tau$, where $\tau$ is a polynomial $\tau$-function of the hierarchy, and $\theta$ is also a polynomial. This fact can be seen as a consequence of the Sato formula for the Baker-Akhiezer function (see below).

The $\Lambda=1$ case corresponds to the KdV hierarchy, where $L=-H=\partial_z^2-u$, with $u=-2(\log\tau)''$. Equation $L\psi=0$ is the Tkachenko equation (\ref{Tkachenko}) for $\theta$ and $\tau$ rewritten in the Schrodinger form:
$$
L\psi=\left(\partial_z^2+2(\log\tau)''\right)[\theta/\tau]=0 \iff \tau''\theta-2\tau'\theta'+\tau\theta''=0 .
$$ 
Here, modulo multiplication by constants, and a shift of $z$, $\theta=P_{n\pm 1}$ and $\tau=P_n$ are the Adler-Moser polynomials. However, the parametrization of the Adler-Moser polynomials in terms of the hierarchy evolution parameters (the ``times") differs from that obtained through Darboux transformations (or polynomial method) in the main body of the paper. Parameters $s_i$ in (\ref{AMExamples}) depend bi-rationally on the KdV times \cite{AM,AcV}. The Adler-Moser polynomials as functions of the KdV times can be expressed as Wronskians of elementary Schur polynomials \cite{Hirota}.

The $\Lambda=2$ case corresponds to the Sawada-Kotera hierarchy (for more details on Sawada-Kotera hierarchy see e.g. \cite{KVdL}). Here, the Lax operator is of the third order $L=\partial_z^3-u\partial_z$, with $u=-6(\log\tau)''$. As we saw in section \ref{3rd_order}, equation $L\psi=0$, $\psi=\theta/\tau$ is equivalent to the $\Lambda=2$ specification of bilinear equation (\ref{bilinear}) for $\tau$ and $\rho=\theta'\tau-\theta\tau'$:
$$
L\psi=\left(\partial_z^3+6(\log \tau)''\partial_z\right)[\theta/\tau]=0  \iff
\left\{\begin{array}{l}\rho''\tau-4\rho'\tau'+4\rho\tau''=0 \\ \rho=\tau'\theta-\tau\theta' \end{array} . \right.
$$
Then, since $\rho$ and $\tau$ are polynomials, polynomial $\tau$-functions of the Sawada-Kotera hierarchy correspond to equilibrium configurations. Here, modulo multiplication by constants, and a shift of $z$, $\theta=q_{n \pm 1}$, and $\tau=q_n$. Zeros of the rational function $\psi'={\rm const}(q_{n\pm 1}/q_n)'$ correspond to positions of charges with ${\cal Q}=-1$, while its poles correspond to positions of charges with ${\cal Q}=2$. The parametrization of $\tau=q_n$ in the times of the Sawada-Kotera hierarchy is different from that of (\ref{pqplus}, \ref{pqminus}) in the integration parameters $s_i$, $r_i$. The polynomials $q_n$ as functions of the Sawada-Kotera times can be presented in the form of Pfaffians.

Below, we will provide explicit expressions for $P_n$ and $q_n$ in terms of the KdV and Sawada-Kotera times, respectively (following \cite{Hirota} and \cite{KVdL}).

First, we recall the definition of elementary Schur polynomials $S_n=S_n(t_1,t_2, \dots, t_n)$:
$$
e^{\sum_{i=1}^\infty t_ik^i}=\sum_{i=0}^\infty S_i k^i ,
$$
$$
S_0=1, \quad S_1=t_1, \quad S_2=t_2+\frac{t_1^2}{2}, \quad S_3=t_3+t_1t_2+\frac{t_1^3}{6}, \quad \dots .
$$
In what follows, $t_i$ will denote evolution parameters (times) of hierarchies, with the first time of the hierarchies identified with $z$:
$$
t_1=z .
$$
For $\Lambda=1$ (KdV hierarchy case) solutions do not depend on the ``even times" $t_2,t_4, \dots $. Up to a normalization factor $3^{n-1}5^{n-2}\cdots(2n-1)$, the Adler-Moser polynomials are expressed through $t_1=z$, and $t_3, t_5, \dots $ as \cite{Hirota}
$$
P_n=\Wr[S_1, S_3, S_5, \dots S_{2n-1}].
$$
Here, all even times $t_{2j}$ are set to zero in all $S_i$, and Wronskian is taken wrt to $t_1=z$. The first several examples of $P_n$ as functions of hierarchy times are:
$$
P_0=1, \, P_1=z, \, P_2=z^3-3t_3, \, P_3=z^6-15z^3t_3+45zt_5-45t_3^2, \, \dots
$$
The dependence of the first several Adler-Moser integration parameters $s_i$ (see (\ref{AMExamples})) on the KdV times $t_3,t_5,t_7,\dots$ is as follows (for more details on bi-rational transformation between parameters, see \cite{AcV}):
$$
s_1=-3t_3, \, s_2=45t_5, \, s_3=-1575t_7, \, s_4=99255\left(t_9-\frac{t_3}{3}\right), \, s_5=9823275\left(\frac{4}{3}t_3^2t_5-t_{11}\right), \, \dots  
$$
The rational Baker-Akhiezer function $\Psi_n(k,z)$ can be obtained using the Sato formula for the KP hierarchy \cite{Z} (since KdV hierarchy is a reduction of the KP hierarchy, where solutions do not depend on the ``even times" $t_{2i}$):
$$
\Psi_n(k,t_1, t_3, t_5, \dots)=\frac{P_n\left(t_1-\frac{1}{k}, t_3-\frac{1}{3k^3}, t_5-\frac{1}{5k^5}, \dots \right)}{P_n(t_1,t_3,t_5,\dots)}e^{kt_1+k^3t_3+k^5t_5+\dots},
$$
$$
L_n\Psi_n(k,z,t_3,t_5, \dots)=k^2\Psi_n(k,z,t_3,t_5, \dots), \quad L_n=\partial_z^2+2(\log  P_n)'' .
$$
The intertwining operator $T_n$ between $L_0=\partial_z^2$ and $L_n=\partial_z^2+2(\log  P_n)''$:
$$
L_nT_n=T_nL_0 ,
$$
can be obtained from the ``rational part" of the Baker-Akhiezer function, by substituting $k \to \partial_z$ (to the right of coefficients) in the following polynomial\footnote{To avoid negative powers of $k$ in $T_n(k)$, the ``rational part" of the Baker-Akhiezer function is multiplied by $k^{\deg P_{n}-\deg P_{n-1}}=k^n$. This ensures that we obtain intertwining operators of minimal order. By multiplying by higher powers of $k$, we obtain intertwining operators of the form $T_n\partial_z^j$.} in $k$ (with rational coefficients in $z$)
$$
T(k)=\frac{P_n\left(t_1-\frac{1}{k}, t_3-\frac{1}{3k^3}, t_5-\frac{1}{5k^5}, \dots \right)}{P_n(t_1,t_3,t_5,\dots)}k^n, \quad t_1=z,
$$
$$
T(k)=k^n+\dots, \quad T(0)={\rm const}\frac{P_{n-1}}{P_n} .
$$
For $\Lambda=2$ (Sawada-Kotera hierarchy case), solutions do not depend on the times $t_{2i}$ and $t_{3i}$, $i>0$. In other words, the polynomial $\tau$-functions depend  on
$$
t_1=z, \,{\rm and} \quad t_5, t_7, t_{11}, t_{13}, t_{17}, t_{19}, \dots 
$$
as well as on an additional set of parameters. Kac and Van de Leur showed \cite{KVdL} that the polynomial $\tau$-functions of the hierarchy can be expressed in Pfaffian form as follows:

Let $\mu=(\mu_1,\mu_2, \dots, \mu_{2m})$, where $m>0$, be either a finite arithmetic progression (cases 1 and 3 below), or a progression extended by $0$ (cases 2 and 4 below): 

\begin{enumerate}

\item For $\tau=q_{2m}$, take sequence $\mu=(6m-2, 6m-5, 6m-8, \dots, 4, 1)$. 

\item For $\tau=q_{2m-1}$, take sequence $\mu=(6m-5, 6m-8, 6m-11, \dots, 4, 1, 0)$.

\item For $\tau=q_{-2m}$, take sequence $\mu=(6m-1, 6m-4, 6m-7, \dots, 5, 2)$. 

\item For $\tau=q_{-2m+1}$, take sequence $\mu=(6m-4, 6m-7, 6m-10, \dots, 5, 2, 0)$.

\end{enumerate}

Also we take sequence\footnote{In this sequence, we set all $c_{2i-1}$ to zero, since they are just shifts of the hierarchy times.} of continuous parameters $c=(0,c_2,0,c_4,0,c_6, \dots, 0, c_{\mu_1+\mu_2-1}, 0) $, in which we substitute recursively (for cases 1-4 respectively):

\begin{enumerate}

\item $c_2=0$, and $c_8, c_{14}, c_{20}, \dots, c_{12m-10}$ ,

\item $c_2=0$, and $c_8, c_{14}, c_{20}, \dots, c_{12m-16}$ ,

\item $c_4, c_{10}, c_{16}, \dots, c_{12m-8}$ ,

\item $c_4, c_{10}, c_{16}, \dots, c_{12m-14}$ ,

\end{enumerate}
by the following formula
$$
c_{2j}=-\frac{1}{2}S_j(2c_2,2c_4,\dots 2c_{2j-2},0), \quad j>1 .
$$
Then, up to multiplication by a constant, the polynomial $\tau$-function is the Pfaffian of the $2m \times 2m$ matrix 
$$
\tau={\rm Pf}\left(\chi_{\mu_i,\mu_j}(\tilde t+c)\right), \quad i,j=1,2,3,\dots, 2m ,
$$
where $\tilde t=(t_1,0,t_3,0,t_5,0,t_7,\dots, 0, t_{\mu_1+\mu_2})=(z,0,t_3,0,t_5,0,t_7,\dots, 0, t_{\mu_1+\mu_2})$, and
$$
\chi_{a,b}(t)=\frac{1}{2}S_a(t)S_b(t)+\sum_{j=1}^b(-1)^jS_{a+j}(t)S_{b-j}(t), \quad a>b\ge 0, \quad \chi_{a,b}=-\chi_{b,a}.
$$
Remark: Since $\tau$ does not depend on $t_{3i}$, $i>0$, to simplify computations, one can set $t_{3i}=0$ and $c_{3i}=0$ to zero in $\tilde t$ and $c$ respectively.

Several first examples of $\tau=q_n$ as functions of $t_i$ and $c_i$ for $n\ge 0$ are
$$
\begin{array}{l}
q_0=1 , \\
q_1=z , \\
q_2=z^5-40c_4z-80t_5 , \\
\begin{aligned}
q_3=z^{12}-440c_4z^8-1760t_5z^7+24640t_7z^5-123200c_4^2z^4-492800t_5c_4z^3 \\
-985600t_5^2z^2+(2956800t_7c_4+1971200t_{11})z-1971200t_7t_5-985600c_4^3 , 
\end{aligned}
\\
q_4=q_4(z,t_5,t_7,t_{11},t_{13},t_{17}; c_4, c_{10}) , \\
q_5=q_5(z,t_5,t_7,t_{11},t_{13},t_{17},t_{19},t_{23}; c_4, c_{10}) , \\
\dots
\end{array}
$$
Here, for $n>1$, $q_n$ is a polynomial in $2(n-1)$ Sawada-Kotera times and $[n/2]$ parameters $c_4,c_{10},\dots, c_{6[n/2]-2}$.

Several first examples of $\tau=q_n$ for $n \le 0$ are
$$
\begin{array}{l}
q_0=1 , \\
q_{-1}=z^2+2c_2 , \\
q_{-2}=z^7+14c_2 z^5+140c_2^2z^3-280t_5z^2-280c_2^3z+1120t_7+1680c_2t_5 , \\
q_{-3}=q_{-3}(z,t_5,t_7,t_{11},t_{13}; c_2, c_8) , \\
q_{-4}=q_{-4}(z,t_5,t_7,t_{11},t_{13},t_{17},t_{19}; c_2, c_8) , \\
q_{-5}=q_{-5}(z,t_5,t_7,t_{11},t_{13},t_{17},t_{19},t_{23},t_{25}; c_2, c_8, c_{14}) , \\
\dots
\end{array}
$$
where $q_{-n}$, $n>0$ is a polynomial in $2n-1$ Sawada-Kotera times and $[(n+1)/2]$ parameters $c_2, c_8, \dots, c_{6[(n+1)/2]-4}$.

Note, that re-parametrization of $q_n$ from $t_i,c_i$ in the above equations to $r_i,s_i$ in eqs. (\ref{pqplus},\ref{pqminus}) is not invertible for $n>3$ or $n<-2$. For instance, $q_4$ in the $t_i,c_i$-parametrization, depends on $z$ and 7 parameters $t_5,t_7,t_{11},t_{13},t_{17}, c_4, c_{10}$. On the other hand, in parametrization (\ref{pqplus}), $q_4$ depends on $z$ and 6 parameters $r_1,r_2,r_3,s_2,s_3,s_4$.

Examples several first $r_i,s_i$, $i>0$ as functions of hierarchy parameters are\footnote{We recall that $r_{\pm i}$ is a coefficient to $z^{\deg{p_{\pm(i-1)}}}$ in polynomial $p_{\pm i}(z)$, while $s_{\pm i}$ is a coefficient to $z^{\deg{q_{\pm(i-1)}}}$ in polynomial $q_{\pm i}(z)$, see (\ref{sr}). Thus $r_i$, $s_i$ depend polynomially on hierarchy parameters.}
$$
\begin{array}{l}
r_1=20 t_{{5}},\\
r_2=-1126400 t_{{11}}-2252800 t_{{7}}c_{{4}},\\
r_3=2022955827200 t_{{17}}+4045911654400 t_{{13}}c_{{4}}-1857816576000 t_{{7}}{t_{{5}}}^{2}+{\frac {41986654617600}{13}} {c_{{4}}}^{3}t_{{5}},\\
r_4={\frac {1790769049177292800000}{13}} t_{{5}}{c_{{4}}}^{2}c_{{10}}-{\frac {87349994569529753600000}{3211}} {t_{{5}}}^{3}{c_{{4}}}^{2}+{\frac {369072026107789312000000}{3211}} {c_{{4}}}^{3}t_{{11}}\\
\begin{array}{r}
-{\frac {175608648299786240000000}{3211}} {c_{{4}}}^{4}t_{{7}}+{\frac {5091682715182694400000}{247}} t_{{5}}c_{{4}}{t_{{7}}}^{2}-63278058274816000000 t_{{13}}c_{{10}}\\
-88864403577241600000 t_{{5}}t_{{11}}t_{{7}}-31639029137408000000 t_{{23}}-63278058274816000000 t_{{19}}c_{{4}} ,
\end{array}
\\
\dots
\end{array}
$$
and
$$
\begin{array}{l}
s_2=-40\,c_{{4}},\\
s_3=24640\,t_{{7}},\\
s_4=-58643200\,c_{{10}}-29321600\,{t_{{5}}}^{2},\\
s_5=350686336000\,t_{{13}}-{\frac {15039048640000}{19}}\,t_{{5}}{c_{{4}}}^{2}, \\
\dots
\end{array}
$$
and so forth.

The rational Baker-Akhieser function can be obtained using the Sato formula for the BKP hierarchy (since the Sawada-Kotera hierarchy is a reduction of the BKP hierarchy, where solutions are independent of $t_{3i}$, see e.g. \cite{KVdL} for details)
$$
\Psi_n(k,t_1,t_5,t_7,\dots)=\frac{q_n\left(t_1-\frac{2}{k},t_5-\frac{2}{5k^5},t_7-\frac{2}{7k^7}, \dots\right)}{q_n(t_1,t_5,t_7, \dots)}e^{kt_1+k^5t_5+k^7t_7+\dots},
$$
$$
L_n\Psi_n(k,z,t_5,t_7,\dots)=k^3\Psi_n(k,z,t_5,t_7,\dots) , \quad L_n=\partial_z^3+6(\log q_n)''\partial_z.
$$
The intertwining operator $T_{\pm n}$ between $L_0=\partial_z^3$ and  $L_{\pm n}=\partial_z^3+6(\log q_{\pm n})''\partial_z$ :
$$
L_{\pm n}T_{\pm n}=T_{\pm n}L_0,
$$
can be obtained from the ``rational part" of the Baker-Akhiezer function, by substituting $k\to\partial_z$ (to the right of coefficients) in the following polynomial in $k$ (with rational in $z$ coefficients)
$$
T_{\pm n}(k)=\frac{q_{\pm n}\left(t_1-\frac{2}{k},t_5-\frac{2}{5k^5},t_7-\frac{2}{7k^7}, \dots\right)}{q_{\pm n}(t_1,t_5,t_7, \dots)}k^{3n-\frac{1}{2}(3\pm 1)}, \quad t_1=z , \quad n>0 ,
$$
$$ 
T_{\pm n}(k)=k^{3n-\frac{1}{2}(3\pm 1)}+\dots, \quad T_{\pm n}(0)={\rm const}\frac{q_{\pm (n-1)}}{q_{\pm n}} .
$$
Examples of the first two intertwining operators for $n>0$, written in the $r_i,s_i$-parametrization (see (\ref{pqplus})) are as follows: The operator
$$
T_1=\partial_z-\frac{2}{z}
$$
intertwines $L_0=\partial_z^3$ with
$$
L_1=\partial_z^3-\frac{6}{z^2}\partial_z .
$$
The operator
$$
T_2=\partial_z^{4}-{\frac { 2\left( 5{z}^{4}+s_{{2}} \right)}{{{z}^{5}-4r_{{1}}+s_{{2}}z}}}\partial_z^{3}+{\frac {40{z}^{3}}{{z}^{5}-4r_{{1}}+s_{{2}}z}}\partial_z^{2}-{\frac {80{z}^{2}}{{z}^{5}-4r_{{1}}+s_{{2}}z}}\partial_z+{\frac {80z}{{z}^{5}-4r_{{1}}+s_{{2}}z}}
$$
intertwines $L_0$ with
$$
L_2=\partial_z^3-6\,{\frac {5\,{z}^{8}+80\,{z}^{3}r_{{1}}-10\,{z}^{4}s_{{2}}+s_2^2}{ \left( {z}^{5}-4\,r_{{1}}+s_{{2}}z\right) ^{2}}}\partial_z ,
$$
and so forth.

Note that the transformation $L_0\to L_n$ can be performed through a chain of permutations and re-factorizations, involving the first-order operators $\partial_z-V_i$, where $V_i$ are rational (see e.g. \cite{BHY,HM}). As we saw in section \ref{3rd_order_equilibrium}, in contrast to the $\Lambda=1$ case, intertwining identities cannot, in general, be constructed solely by permuting factors in $L_n$: The factorization chain now includes intermediate iterations. For instance, for $n>0$, the order of the intertwining operator $T_n$ is $3n-2$, i.e., ${\rm ord}(T_{n+1})-{\rm ord}(T_n)=3$. This is because each transformation $L_n\to L_{n+1}$, $n>0$, presented by (\ref{DTu3},\ref{DTkappa}), actually corresponds to three steps, i.e., to three permutations and re-factorizations in the chain ($L_0 \to L_1$ is performed in a single step):
$$
L_n=B_nA_n \to A_nB_n=\tilde B_n\tilde A_n \to \tilde A_n \tilde B_n = \dbtilde{B}_n\dbtilde{A}_n \to \dbtilde{A}_n\dbtilde{B}_n=B_{n+1}A_{n+1}=L_{n+1} , \quad n>0 . 
$$
As a consequence, the operators ${\cal T}_n=\dbtilde{A}_n\tilde A_n A_n$, $n>0$ that intertwine $L_n$ and $L_{n+1}$:
$$
{\cal T}_nL_{n}=L_{n+1}{\cal T}_n, \quad T_n={\cal T}_{n-1}{\cal T}_{n-1}\cdots {\cal T}_1{\cal T}_0, \quad {\cal T}_n=\partial_z^3+\dots, \quad {\cal T}_0=T_1=\partial_z-\frac{2}{z},
$$
are of the third order, when $n>0$ (of the first order when $n=0$, ${\cal T}_0=A_0$). For example
$$
{\cal T}_1=\partial_z^{3}-{\frac { 8\left( r_{{1}}+{z}^{5} \right)}{ \left( {z
}^{5}-4\,r_{{1}}+s_{{2}}z \right) z}}\partial_z^2+2\,{\frac { 9\,{z}^{5}-3\,s_{{2}}z+
4\,r_{{1}} }{ \left( {z}^{5}-4\,r_{{1}}+s_{{2}}z
 \right) {z}^{2}}}\partial_z, \quad {\cal T}_1L_1=L_2{\cal T}_1 .
$$
Closing this section, we note that (as follows from \cite{W}), similarly to the KdV case, operators $L_n$ are bi-spectral.

\end{appendices}

%%%%%%%%%%%%%%%%%%%%%%%%%%%%%%%%%%%%%%%%%%%%%%%

\end{document}